\documentclass[11pt]{article}
\usepackage[margin=1in]{geometry}
\usepackage{lmodern}
\usepackage[T1]{fontenc}
\usepackage[utf8]{inputenc}
\usepackage{microtype}
\usepackage{amsmath,amssymb}
\usepackage[dvipsnames]{xcolor}
\usepackage{graphicx}
\usepackage{booktabs}
\usepackage{tabularx}
\usepackage{multirow}
\usepackage{enumitem}
\usepackage{caption}
\usepackage{algorithm}
\usepackage[noend]{algpseudocode}
\usepackage[square]{natbib}
\usepackage[colorlinks=true,linkcolor=blue!60!black,citecolor=blue!60!black,urlcolor=blue!60!black]{hyperref}

\newcommand{\Regret}{\mathrm{Regret}}
\newcommand{\Cov}{\mathrm{Cov}}
\newcommand{\Var}{\mathrm{Var}}
\newcommand{\E}{\mathbb{E}}
\newcommand{\tr}{\operatorname{tr}}
\DeclareMathOperator{\clip}{clip}

\title{AI Governance for Institutional Readiness in Finance}
\author{Irene Aldridge\thanks{Irene@RiskAICenter.com, Irene@AbleAlpha.com}
\and Steve Krawciw\thanks{Steve@RiskAICenter.com, Steve@AbleMarkets.com}\\[4pt]}

\date{November 2026}

\begin{document}
\maketitle

\begin{abstract}
\noindent Agentic AI is gaining acceptance in asset management, but governance has not kept pace: 88\% of surveyed finance professionals report no operational governance framework for agentic AI, and only 24 of 75 large U.S.\ money managers disclosing AI use in Form ADV filings report a formal governance policy. We argue this gap is architectural: governance built for static validation does not survive continuously retrained agentic policies. We propose a four-layer framework (Policy, Engineering, Composition, Systemic) grounded in two distinct kinds of evidence, kept explicitly separate: two calibrated synthetic illustrations (an integrated-covariance regret monitor for policy drift; a crowding simulation showing joint drawdown risk rising from 39.2\% to 79.3\%), and three real, documented cases (a deployed LLM-embedding trading strategy, a \$45 billion discretionary fund's forced-deleveraging blowup, and a tribunal ruling holding an airline liable for its chatbot). The drift monitor rests on an exact identity: regret equals the covariance between costs and decisions, integrated along the path from certainty to realized uncertainty. The plain covariance, computable from observed costs and decisions alone, is therefore always an upper bound on regret; it equals exactly twice the regret for affine decision layers such as budget-constrained mean--variance portfolio construction; and replay access to the decision layer, without access to model weights, makes the monitor exact. The synthetic examples demonstrate computability from observable data; the cases demonstrate that the failure modes are not hypothetical. We provide a 90-day implementation sequence spanning trading and payments/customer-facing systems.
\end{abstract}


\vspace{1em}

\tableofcontents
\section*{Data availability statement}
The data and code used to produce the worked numerical examples in Sections~\ref{sec:l2drift}, \ref{sec:l2fraud} and~\ref{sec:l3calib} and in Appendix~\ref{app:robust} (the two integrated-covariance drift monitors, their robustness grids, and the two-agent crowding simulation) are fully synthetic, generated by the code itself; the code is provided as supplementary material. No proprietary or confidential data were used in these experiments. The informal survey data described in Section~\ref{sec:survey} were collected via a self-selected LinkedIn poll and are summarized in aggregate only; no individual-level responses were retained. The Form ADV filing data referenced in Section~\ref{sec:adv} are publicly available from the U.S.\ Securities and Exchange Commission's Investment Adviser Public Disclosure database (\url{https://adviserinfo.sec.gov}).

\section*{Declaration of competing interest}
The authors declare that they have no known competing financial interests or personal relationships that could have appeared to influence the work reported in this paper.

\section{Introduction: Awareness Without Action as a Systemic Risk}\label{sec:intro}
Consider the following hypothetical scenario: a hedge fund desk deployed an LLM-embedding trading signal eighteen months ago. The signal has performed well: a Sharpe ratio north of 2, consistent with the published literature on this class of strategy. The model retrains monthly on an expanding window. It runs against a licensed third-party embedding model that the hedge fund did not build and does not control. No one on the team has looked at the model's internal weights in over a year.

Three questions follow, and this paper is organized around answering all three:
\begin{enumerate}
\item \textbf{What does the hedge fund need to monitor, and how often?} The model's live behavior six months after deployment is not guaranteed to resemble the behavior the team validated at launch. Furthermore, unlike a rules-based execution algorithm, there is no fixed decision logic to check against.
\item \textbf{Who is accountable if the vendor changes the underlying embedding model?} A vendor-side model upgrade can shift the strategy's return distribution without tripping any single alert that the hedge fund currently has in place. This can happen because no single component of the pipeline, such as the data feed, the embedding model, or the portfolio construction step, is individually broken.
\item \textbf{What happens if other desks or hedge funds are running the same trade?} If competitors are training similar models on the same public news feed towards similar objectives, the strategy's true risk may depend on decisions being made on unknown desks that one cannot see, in ways that the fund's own risk process cannot detect.
\end{enumerate}

Most governance frameworks currently in use at asset managers were not built to answer these questions. They were built for algorithmic trading systems with fixed, auditable decision rules. As such, the governance frameworks incorporate rigid pre-, intra-, and post-trade controls that assume the model behaves consistently between validation cycles. An agentic or continuously-retrained system violates that assumption by design. This is not a gap in awareness; our survey evidence below confirms that nearly every finance professional we polled already knows that agentic AI is in production somewhere in their firm. It is a gap in \emph{what to build}: a specific, implementable set of monitoring tools, escalation triggers, and disclosure practices matched to how these systems actually behave in production.

This paper proposes four such tools, organized as layers a firm can adopt incrementally rather than all at once:
\begin{itemize}
\item A \textbf{policy layer} treating the reward function driving an agentic strategy with the same review rigor as any other risk policy.
\item An \textbf{engineering layer} built around two concrete, computable alarms: an exact measure of how much a live strategy's objective depends on cost information, based on the identity that regret equals integrated cost--decision covariance and tracked against its validated level (Section~\ref{sec:l2}), and a kill-switch tied to model-internal confidence signals rather than a model's self-reported confidence.
\item A \textbf{composition layer} for pipelines, like the embedding-based strategy above, that depends on third-party components no single audit can fully see into.
\item A \textbf{systemic layer} for the crowding risk that emerges only when a unit's governance process is compared against everyone else's.
\end{itemize}

Survey evidence shows that finance professionals are almost universally aware of agentic AI deployment, yet a large majority report having no operational governance process for it. This is not the profile of a technology that institutions have failed to notice; it is the profile of a technology whose governance requirements are not yet well understood enough to act on.

Agentic systems, particularly those incorporating reinforcement learning or continuously retrained statistical learning, update their decision policies in production. The longer a firm operates without governance, the further the live system drifts from whatever baseline was validated at deployment. This paper (i) quantifies the governance gap with survey and regulatory evidence; (ii) identifies, formally, four failure modes that defeat governance frameworks built for deterministic systems; (iii) proposes a layered governance architecture; and (iv) grounds Layers 3 and 4 in a concrete, currently-deployed class of strategy: LLM-embedding-based news trading, as documented empirically by \citet{didisheim2026}.

For each layer, we present a concrete, currently-deployed example, the LLM-embedding news-trading strategy sketched above, documented empirically by \citet{didisheim2026}. Section~\ref{sec:gap} quantifies how wide the awareness-to-governance gap actually is, using both survey evidence and Form ADV disclosure data. Sections~\ref{sec:fail}--\ref{sec:framework} lay out why existing frameworks fail and what should replace them. Section~\ref{sec:roadmap} gives a 90-day implementation sequence. A reader managing the strategy described above should be able to leave this paper with a specific answer to each of the three questions posed here.

\section{Related Work}\label{sec:related}
This paper draws on and departs from four largely separate literature streams:
\begin{itemize}
\item traditional model risk management in finance;
\item market microstructure and systemic crowding risk;
\item AI safety research on reward hacking and policy drift; and
\item the emerging body of regulatory guidance on AI in capital markets.
\end{itemize}
Each stream addresses part of the problem posed in Section~\ref{sec:intro}; none addresses all of it jointly, which is the gap that this paper's four-layer framework is designed to close. The drift monitor of Layer~2 additionally draws on the value-of-information and predict-then-optimize literatures (Section~\ref{sec:voi}).

\subsection{Traditional Model Risk Management in Finance}\label{sec:mrm}
Model risk management, as currently practiced, traces back to supervisory guidance issued in the wake of the 2007--2009 financial crisis, principally SR~11-7 \citep{sr117} and its international analogs. This approach requires independent validation of a model's conceptual soundness, ongoing monitoring, and outcomes analysis. The regime assumes, largely implicitly, that a validated model produces stable, auditable outputs between periodic review cycles. In Section~\ref{sec:fail}, we argue that this assumption is structurally violated by continuously or periodically-retrained agentic systems. Basel Committee guidance on operational and model risk extends this logic to a broader set of quantitative risk models but shares the same static-validation premise. Pre-, intra-, and post-trade rules engines, the dominant governance technology for algorithmic execution over the past two decades, sit within this same tradition: they constrain what a system is permitted to do, not whether the policy generating its decisions has drifted from what was validated.

\subsection{Market Microstructure, Real-Time Risk, and Systemic Crowding}\label{sec:micro}
A separate literature examines how automated and high-frequency trading systems fail in production, independent of any formal governance question. \citet{aldridgekrawciw2017} argues that the compression of decision and execution into microsecond timeframes has outpaced traditional, periodic risk-management cycles and calls for risk monitoring architected to operate at the same time scale as the trading systems it governs. As such, \citet{aldridgekrawciw2017} is a precursor, in spirit, to this paper's Layer~2 requirement that policy-stability monitoring be continuous and real-time rather than being reviewed on a fixed audit calendar. \citet{kirilenko2017} provide the canonical empirical account of a single-day systemic event (the May 6, 2010 Flash Crash) driven by the interaction of automated trading strategies under stress, showing that intermediaries did not change their behavior as prices fell. This finding is directly relevant to this paper's Layer~2 argument that automated systems can behave exactly as designed at the component level while producing a systemically destabilizing aggregate outcome. \citet{khandanilo2007} document the closest empirical precedent to this paper's Layer~4 crowding mechanism: the August 2007 ``Quant Meltdown,'' in which the forced deleveraging of one or more large quantitative equity portfolios propagated losses across many independently-managed, similarly-constructed long/short equity funds that had no formal connection to one another, through a purely market-mediated (price-impact) channel rather than any shared governance failure. This is the pre-AI, human-strategy analog of the mechanism Section~\ref{sec:l3} formalizes for agentic strategies trained on shared public data.

A newer theoretical and simulation literature studies what happens when the strategies themselves are reinforcement-learning agents. \citet{dou2025collusion} show that informed speculators trading with RL algorithms autonomously sustain collusive, supra-competitive profits without agreement, communication, or intent, through price-trigger strategies or through an over-pruning bias in learning; in both cases the agents under-react to their private information, and price informativeness and liquidity deteriorate. \citet{dou2026instability} extend the setting to bubbles: RL arbitrageurs with noisy signals may either fail to coordinate, which stabilizes prices, or learn to ride bubbles collectively, which destabilizes them, depending on the market environment. \citet{dou2026planning} studies agentic RL speculators with explicit planning modules and find that, when feedback trading is strong, they converge on intertemporal strategies that create and exploit bubbles and amplify market fragility. For governance, this literature contributes two points the human-strategy precedents cannot. Harmful coordination can arise among agents that no one instructed to coordinate, which is invisible to rules engines and to any review of intent (Section~\ref{sec:rules}). And correlated behavior across institutions can be strategic rather than only the mechanical consequence of shared inputs, a second channel for the cross-institutional risk of Section~\ref{sec:l3}.

\subsection{AI Safety: Reward Hacking and Policy Drift}\label{sec:aisafety}
A distinct literature, originating outside finance, formalizes the failure mode that this paper's Layer~1 and Layer~2 are built to govern. \citet{amodei2016} frame reward hacking as an agent finding an unintended way to maximize a proxy objective that diverges from the designer's true intent. Reward hacking is one of a small set of concrete, near-term AI safety problems requiring engineering solutions rather than philosophical ones. \citet{skalse2022} formalize this concern, providing a precise definition of when a proxy reward function is ``hackable'' relative to a true objective and showing that, absent restrictive assumptions, most proxy/true reward pairs are hackable in principle. This paper's Layer~1 requirement that a reward function receive the same review rigor as any other risk policy prior to deployment operationalizes this literature's central finding for a financial setting. An unreviewed reward function is not a training-time technicality but a governance object whose specification determines whether the deployed system's incentives are aligned with the institution's actual risk tolerance. This literature has, to date, developed largely independently of financial model risk management (Section~\ref{sec:mrm}); one contribution of this paper is to translate its concepts (proxy objectives, policy drift, hackability) into the vocabulary and institutional structures (SR~11-7-style review, kill-switch architecture) within which finance risk functions already operate.

\subsection{Algorithmic Fair-Lending and Disparate-Impact Literature}\label{sec:fairlending}
Section~\ref{sec:aisafety} situates this paper's Layer~1 and Layer~2 in the AI-safety reward-hacking literature, following \citet{amodei2016} and \citet{skalse2022}. That literature formalizes when a proxy objective diverges from a designer's true intent, but neither paper addresses the specific legal and empirical form that divergence takes in consumer lending and, by direct extension, fraud decision-making: disparate impact on a protected class. The fair-lending exposure this paper's Section~\ref{sec:l1fraud} identifies in an unreviewed cost matrix is not a novel observation about algorithmic credit and fraud systems; it is a well-established empirical finding in a distinct literature, and grounding it there is necessary before Section~\ref{sec:l1fraud}'s claim can be read as more than an analogy borrowed from the reward-hacking literature.

\citet{barocas2016} provide the foundational legal-and-technical account of how a facially neutral, accuracy-optimizing statistical model can produce a disparate impact on a protected class without any protected characteristic appearing as a model input, precisely because proxies for the characteristic remain predictive of the outcome the model is trained to minimize. This is the general mechanism Section~\ref{sec:l1fraud} restates for the specific case of a fraud cost matrix: a model tuned to minimize $c_{FN}$ without a reviewed constraint on $c_{FP}$ can learn to decline disproportionately along exactly the kind of proxy channel \citet{barocas2016} formalize, without the model or a reviewer inspecting only its accuracy metrics ever encoding the protected characteristic directly.

Two empirical papers in consumer finance confirm that this is not merely a theoretical risk in this asset class. \citet{fuster2022} show, using U.S.\ mortgage data, that replacing a traditional underwriting model with a machine-learning model increases the disparity in credit outcomes between and within racial and ethnic groups, and that this disparity is attributable primarily to the additional flexibility of the ML model and its ability to uncover statistical relationships that a simpler model would not, rather than to any input resembling a protected characteristic. \citet{bartlett2022} show, using an identification strategy built on GSE and FHA mortgage-pricing rules, that algorithmic (``FinTech'') lenders exhibit measurably lower, but still statistically significant and economically material, rate disparities against Latinx and Black borrowers relative to traditional lenders. This is evidence that algorithmic underwriting reduces but does not eliminate the exposure Section~\ref{sec:l1fraud} treats as a mandatory Layer~1 review item.

\paragraph{Why this literature, and not only the AI-safety literature, grounds Section~\ref{sec:l1fraud}.} Section~\ref{sec:aisafety} argues that this paper's contribution is translating concepts from the AI-safety literature (proxy objectives, policy drift, hackability) into financial governance vocabulary. The fair-lending literature surveyed here does the complementary work of establishing that, in consumer credit specifically, the proxy-objective failure mode is not a hypothetical construction from first principles but a repeatedly measured empirical regularity, predating agentic AI by at least a decade and persisting, in attenuated form, even in the algorithmic lenders \citet{bartlett2022} study. Section~\ref{sec:l1fraud}'s disparate impact analysis requirement is best read as extending this literature's established empirical finding (machine-learning flexibility predictably produces disparate impact absent a reviewed constraint) from underwriting to the adjacent domain of real-time fraud decisioning, rather than as an independent claim resting only on the AI-safety literature's more general and less finance-specific reward-hacking formalism.

\subsection{Regulatory and Policy Frameworks for AI in Finance}\label{sec:regframe}
Regulators and standard-setters have begun to address agentic and AI-driven financial systems directly, though largely at the level of principles rather than implementable controls. IOSCO's 2021 report on AI and machine learning use by market intermediaries and asset managers \citep{iosco2021} and its 2025 follow-up on AI in capital markets \citep{iosco2025}, the latter explicitly extending its scope to generative and agentic AI, both identify governance gaps consistent with this paper's survey evidence (Section~\ref{sec:survey}), but stop short of proposing a layered, implementable architecture. The EU AI Act \citep{euaiact2024} introduces a risk-tiered regulatory structure that classifies AI systems by application domain and potential for harm, offering a coarse-grained analog to this paper's Agentic Risk Score (Section~\ref{sec:taxonomy}), but calibrated for cross-sectoral regulation rather than for the specific dynamics of continuously-retrained financial strategies. The NIST AI Risk Management Framework \citep{nist2023} proposes a general-purpose govern/map/measure/manage lifecycle applicable across industries. This framework is a natural complement to, rather than a substitute for, the finance-specific mechanisms (integrated-covariance regret monitoring, vendor-version attestation, policy-similarity disclosure) this paper develops in Sections~\ref{sec:l2}--\ref{sec:l3}, since none of these general frameworks specifies an exact, computable statistic for detecting policy drift in a deployed financial strategy from its observed costs and decisions, without access to the model's internal state.

\subsection{Regulatory and Policy Frameworks for AI in Finance, Extended: Payments-Specific Regulation}\label{sec:payreg}
Section~\ref{sec:regframe} surveys AI-specific regulatory guidance (IOSCO, the EU AI Act, NIST AI RMF) that addresses agentic systems at the level of general principle rather than domain-specific, implementable controls. The extensions proposed in Sections~\ref{sec:l2kill}--\ref{sec:l3fraud} sit inside a second, older, and considerably more prescriptive regulatory layer that Section~\ref{sec:regframe} does not address, because it was not written with AI in mind at all: payments-specific consumer-protection and financial-crime regulation that predates agentic systems but binds their behavior regardless. This layer matters for the paper's argument in a way the AI-specific frameworks of Section~\ref{sec:regframe} do not, because it is the one layer among those discussed in this paper that imposes a hard, external clock on remediation, a constraint absent from every other governance dimension this paper formalizes.

\paragraph{Regulation E and the error-resolution clock.} In the United States, Regulation E (implementing the Electronic Fund Transfer Act) imposes fixed timelines within which a financial institution must investigate and resolve a consumer's assertion of an unauthorized or erroneous electronic funds transfer, typically requiring a preliminary determination within ten business days (extendable under defined conditions) and a final resolution within 45 days. This is directly relevant to the bounded action authority control of Section~\ref{sec:l2bounded}: an agent's decline or hold decision that a customer disputes does not merely trigger an internal governance review on the institution's own schedule, as an internal drift alarm under Section~\ref{sec:l2} would; it starts a regulatory countdown independent of whether the institution's Layer~2 monitoring has yet flagged the underlying decision as anomalous. A kill-switch or bounded-action-ceiling breach detected on day 30 of a Reg~E investigation window is governance operating correctly but too late to avoid a compliance deadline, a failure mode with no analog in the trading context of Sections~\ref{sec:l2}--\ref{sec:l25}, where no comparable statutory clock applies to a drifted position.

\paragraph{BSA/AML and suspicious activity reporting.} The Bank Secrecy Act and its implementing regulations require covered institutions to file a Suspicious Activity Report (SAR) within 30 calendar days of detecting facts that may constitute a basis for filing (extendable to 60 days absent an identified suspect), and separately prohibit disclosing the existence of a SAR to the subject of the report. This creates a governance requirement with no counterpart elsewhere in this paper's framework: a fraud-decision-making agent's outputs and internal reasoning traces may themselves constitute part of the evidentiary basis for a SAR filing, meaning the audit-logging and vendor-attestation requirements of Section~\ref{sec:l25fraud} must be designed to preserve a BSA/AML-compliant chain of evidence, including the confidentiality restriction on SAR disclosure, which places a specific constraint on how Layer~3's joint-output sampling and vendor-attestation records (Section~\ref{sec:l25}, item~5) may be stored, accessed, and by whom; this has no equivalent constraint in the trading-governance literature Section~\ref{sec:mrm} surveys.

\paragraph{PSD2/PSD3 and cross-border fraud-liability shifting.} In the European Union and the United Kingdom, the second and forthcoming third Payment Services Directives (PSD2 and the proposed PSD3/Payment Services Regulation) establish strong customer authentication requirements and, more consequentially for this paper's crowding argument, allocate fraud liability between payment service providers based in part on each party's authentication and fraud-control practices. This is directly relevant to Section~\ref{sec:l3fraud}'s consortium-crowding mechanism: where a shared consortium signal or a concentrated upstream fraud-scoring vendor (Section~\ref{sec:l3fraud}, mechanism~2) produces a correlated miscalibration event across institutions, PSD2/PSD3's liability-allocation framework determines which participating institution bears the resulting loss, a legal consequence of crowding that Section~\ref{sec:l3}'s trading-crowding formalization does not need to resolve, since no comparable inter-institutional liability-shifting regime governs correlated drawdowns among independently-managed trading strategies.

\paragraph{Positioning relative to Table~\ref{tab:coverage}, and a second table for this layer.} These three regimes do not fit the AI governance-maturity axis Table~\ref{tab:coverage} scores (Policy-layer objective review, Engineering-layer real-time monitoring, Composition-layer risk, Systemic-layer risk) because none were designed with AI governance as their object at all; Reg~E, BSA/AML, and PSD2/PSD3 govern the underlying financial activity (a disputed transfer, a suspicious transaction, a cross-border payment) irrespective of whether an agentic system, a rules engine, or a human analyst produced the decision. We therefore do not add a row to Table~\ref{tab:coverage}, since these are not competing or complementary AI-governance frameworks in the sense that the table catalogs. Table~\ref{tab:payreg} instead positions them on a different axis entirely---not maturity of AI governance coverage, but which of this paper's layers each regime binds, and by what mechanism, since this is the dimension a firm actually needs when designing escalation paths: a fixed regulatory substrate that any Layer~2 or Layer~3 control proposed for a fraud or customer-facing pipeline in Sections~\ref{sec:l2kill}--\ref{sec:l3fraud} must be designed to satisfy as a binding external constraint, not as an optional best practice competing for adoption priority the way the frameworks in Table~\ref{tab:coverage} are. A firm that implements this paper's extended Layer~2 kill-switch (Section~\ref{sec:l2kill}) and bounded-action-authority control (Section~\ref{sec:l2bounded}) without designing their escalation and audit-logging paths around Reg~E's and BSA/AML's fixed clocks, as Table~\ref{tab:payreg} makes explicit, has built a technically sound monitor that will nonetheless fail a compliance deadline the monitor itself was never designed to track.

\begin{table}[ht]
\centering\small
\caption{Payments-specific regulatory constraints binding this paper's Layer~2 and Layer~3 controls}\label{tab:payreg}
\begin{tabularx}{\textwidth}{>{\raggedright\arraybackslash}p{2.3cm}>{\raggedright\arraybackslash}X>{\raggedright\arraybackslash}p{2.6cm}>{\raggedright\arraybackslash}X}
\toprule
Regime & Binding clock / trigger & Layer(s) constrained & Nature of the constraint\\
\midrule
Regulation E (EFTA) & 10-business-day preliminary determination; 45-day final resolution, from date of customer dispute & Layer 2 (\S\ref{sec:l2kill}), Layer 2 extended (\S\ref{sec:l2bounded}) & Statutory investigation-and-resolution deadline; independent of whether internal drift or ceiling-breach monitoring has already flagged the underlying decision\\
\addlinespace
BSA/AML (SAR filing) & 30 calendar days from detection of reportable facts (extendable to 60 days absent an identified suspect) & Layer 3 (\S\ref{sec:l25fraud}) & Evidentiary-chain and filing deadline, coupled with a statutory non-disclosure requirement (SAR confidentiality) that constrains where and how audit logs may be stored and who may access them\\
\addlinespace
PSD2 / PSD3 (proposed PSR) & Triggered at the point of a cross-border fraud loss event; no fixed calendar window, but liability allocation is determined at that point & Layer 4 (\S\ref{sec:l3fraud}) & Inter-institutional liability-allocation rule, based in part on each party's authentication and fraud-control practices at the time of the loss\\
\bottomrule
\end{tabularx}
\end{table}

\subsection{LLM-Native Financial Strategies}\label{sec:llmfin}
A recent and fast-growing literature documents financial strategies whose signal generation depends directly on large language models, rather than on LLMs as a research aid. \citet{didisheim2026} construct and empirically validate the LLM-embedding news-trading strategy (MSRR) used as this paper's Layer~3/4 case study (Section~\ref{sec:msrr}), and \citet{he2025} establish the chronological-consistency methodology. This methodology ensures that an LLM's apparent predictive power is not an artifact of training-data lookahead. \citet{didisheim2026} apply this methodology, and we use it as well as evidence that vendor model-version attestation (Section~\ref{sec:l25msrr}) is necessary but not sufficient for Layer~3 due diligence. \citet{chen2026} develop the inner-confidence quantification method we use in Layer~2 kill-switch architecture (Section~\ref{sec:l2}). The methodology addresses a problem specific to LLM-based decision systems: that declared model confidence is unreliable. The methodology has no analog in traditional model risk management, since a linear regression or a decision tree has no equivalent ``declared confidence'' to distrust.

\subsection{Value of Information and Predict-then-Optimize}\label{sec:voi}
The Layer~2 drift monitor measures regret: the loss from acting on a cost estimate rather than on the realized cost. In stochastic programming this is the expected value of perfect information \citep{raiffa1961,birge2011}; in machine learning it is the regret of predict-then-optimize pipelines, in which decisions are made from predicted costs \citep{elmachtoub2022}. Both are usually estimated by re-solving the decision problem scenario by scenario, which reports how large regret is but not what generates it. \citet{aldridge2026} show that regret is generated by the co-movement of costs and optimal decisions, and make this exact: regret equals the cost--decision covariance integrated along the path from the mean to each scenario, the plain covariance is always an upper bound, and for affine decision rules, as in budget-constrained mean--variance portfolio construction, regret is exactly half the covariance. The identity holds scenario by scenario for any predicted cost, including for linear and combinatorial decision rules such as an approve/decline threshold. We use it as the computational core of Layer~2 (Section~\ref{sec:l2}): it turns ``has the live policy drifted from validated behavior?'' into a question about observed costs and decisions, answered without access to the policy's internal state.

\subsection{Positioning of This Paper}\label{sec:position}
Table~\ref{tab:coverage} summarizes the coverage gap that this paper's four-layer framework is designed to close: each existing literature or framework addresses one or two of the four governance dimensions this paper identifies (policy-level objective review, engineering-level real-time monitoring, cross-component composition risk, and cross-institutional systemic risk), but none addresses all four jointly, and none provides a computable, finance-specific instantiation of each layer.

\begin{table}[ht]
\centering\small
\caption{Coverage of existing frameworks and literature against this paper's four governance layers}\label{tab:coverage}
\begin{tabularx}{\textwidth}{>{\raggedright\arraybackslash}Xcccc}
\toprule
Framework / literature & \shortstack{Layer 1\\(Policy)} & \shortstack{Layer 2\\(Engineering)} & \shortstack{Layer 3\\(Composition)} & \shortstack{Layer 4\\(Systemic)}\\
\midrule
SR 11-7 / Basel model risk management & partial & partial & no & no\\
Pre/intra/post-trade rules engines & no & partial & no & no\\
Real-Time Risk \citep{aldridgekrawciw2017} & no & partial & no & no\\
Flash Crash / crowding literature \citep{kirilenko2017,khandanilo2007} & no & no & no & \shortstack{yes (empirical,\\non-AI)}\\
AI-powered trading theory \citep{dou2025collusion,dou2026instability,dou2026planning} & no & no & no & \shortstack{yes (theory and\\simulation)}\\
AI safety / reward hacking \citep{amodei2016,skalse2022} & yes & partial & no & no\\
EU AI Act \citep{euaiact2024} & partial & no & no & no\\
NIST AI RMF \citep{nist2023} & partial & partial & partial & no\\
IOSCO AI reports \citep{iosco2021,iosco2025} & partial & partial & partial & partial\\
This paper & yes & yes & yes & yes\\
\bottomrule
\end{tabularx}
\end{table}

The model risk management literature (Section~\ref{sec:mrm}) supplies the institutional review discipline that Layer~1 borrows, but does not provide a mechanism for real-time drift detection. The market microstructure literature (Section~\ref{sec:micro}) documents crowding and systemic propagation empirically, but focuses on human-strategy portfolios rather than agentic systems; its newer AI-powered-trading branch shows, in theory and simulation, how RL and planning agents can coordinate without intent, but is addressed to market regulators and antitrust rather than to the deploying institution's own governance process. The AI safety literature (Section~\ref{sec:aisafety}) formalizes the drift and reward-hacking mechanisms that Layer~1 and Layer~2 are designed to catch, but it was not developed with financial regulatory structures (SR~11-7-style review, kill-switch architecture tied to trading limits) in mind. The regulatory frameworks (Section~\ref{sec:regframe}) identify the governance gap at the level of principle, consistent with this paper's own survey evidence, but do not provide implementable, computable controls. The LLM-native finance literature (Section~\ref{sec:llmfin}) supplies the empirical ground truth, which is an economically large, currently deployed strategy class. This is the ground truth against which this paper validates Layers~3 and~4, but it was not written as a governance contribution. The value-of-information literature (Section~\ref{sec:voi}) supplies the exact identity on which Layer~2's drift monitor is built, but was not written with governance in mind. This paper's four-layer framework is, to our knowledge, the first to unify these strands into a single, computable, finance-specific architecture spanning reward-function review, real-time drift monitoring, cross-vendor composition risk, and cross-institutional crowding disclosure.

\section{Evidence of the Governance Gap}\label{sec:gap}
\subsection{Survey Evidence}\label{sec:survey}
An informal LinkedIn poll of finance professionals (May 2026) found that 88\% report no operational governance framework for agentic AI, while 0\% report being unaware of agentic AI deployment in their institution or industry. The self-selected sample plausibly skews toward AI-attentive respondents, so the true gap across the industry is likely wider, not narrower, than 88\%.

\subsection{Regulatory Disclosure Evidence}\label{sec:adv}
Of 100 of the largest U.S.\ money managers examined via Form ADV filings, 75 disclosed some form of AI use; of those, only 24 (32\%) disclosed an accompanying formal governance policy.

\subsection{What ``Governance'' Currently Means in Practice}\label{sec:practice}
Pre-, intra-, and post-trade rules engines have governed algorithmic trading for roughly two decades. This is rigorous governance of deterministic systems. The current paper discusses the governance of agentic systems. The behavior of agentic systems six months after deployment may not be fully described by the rules that were appropriate at the time of deployment.

\section{Why Existing Governance Frameworks Fail for Agentic Systems}\label{sec:fail}
\subsection{The Static Assumption}\label{sec:static}
Traditional model risk management (SR~11-7 and successors) assumes a validated model produces consistent outputs for consistent inputs between validation cycles. An RL agent, or any continuously or periodically-retrained policy, violates this by design.

\subsection{VaR and the Policy Drift Problem}\label{sec:var}
VaR-style frameworks bound the loss distribution of systems with fixed decision logic. When the decision logic is not fixed, this is a structural mismatch, not a parameter-estimation problem.

\subsection{Rules Engines Are Necessary but Not Sufficient}\label{sec:rules}
Pre/intra/post-trade controls remain the hard outer boundary on agent behavior, but they do not evaluate intent. Reward hacking operates entirely within rules-engine bounds.

\subsection{The Compliance-as-PDF Problem}\label{sec:pdf}
Governance is an engineering property, not a compliance document: nothing in an agentic system's execution path enforces a policy document the moment the underlying policy updates.

\subsection{Policy Drift Under Adversarial Input Distributions}\label{sec:adversarial}
A model's behavioral guardrails, calibrated against a standard input distribution, can fail once inputs are drawn from an adversarial distribution. We propose an experimental protocol (bypass rate by perturbation type, with confidence intervals) for Layer~2 certification.

\subsection{Compositional Risk: Emergent Failure Across Validated Components}\label{sec:composition}
Even when every individual component of a multi-module pipeline is validated and behaves exactly as specified, the composition of components can produce an outcome that no component-level audit would surface because the unit of analysis that fails is the pipeline, not any single model.

\section{A Four-Layer Framework for Agentic AI Governance}\label{sec:framework}
\subsection{Layer 1: Policy Layer (What You Intend)}\label{sec:l1}
The reward function is the mathematical expression of institutional intent and should receive the same review, approval, and version control as any other risk policy document.

\subsubsection{Layer 1, Extended: The Reward Function as a Cost-Matrix Review for Fraud-decision-making Agents}\label{sec:l1fraud}
Section~\ref{sec:l1} establishes that the reward function is the mathematical expression of institutional intent and should receive the same review, approval, and version control as any other risk policy document. For a trading agent, the reward function is typically a risk-adjusted return objective, and Section~\ref{sec:sal1} shows what happens when an analogous governing objective, a discretionary investment thesis, is never subjected to this review. For a fraud-decision-making agent, the object that plays the role of the reward function is the false-positive/false-negative cost matrix that the model is trained or tuned to minimize. We propose that this cost matrix, rather than the model architecture, the feature set, or the aggregate accuracy metric an institution typically reviews, is the correct object of Layer~1 review for fraud pipelines, and that treating it as such surfaces a fair-lending exposure that model-accuracy review alone does not.

\paragraph{The fraud-domain reward function.} A fraud-decision-making agent is trained, explicitly or implicitly, against a cost matrix of the form shown in Table~\ref{tab:costmatrix}, where $c_{FN}$ is the realized cost of a false negative (the fraud loss, chargeback liability, and downstream investigation cost of an approved fraudulent transaction) and $c_{FP}$ is the realized cost of a false positive (a declined legitimate transaction). In practice $c_{FP}$ is rarely priced with the same rigor as $c_{FN}$: fraud loss is a hard, immediately observable dollar figure that appears directly on a loss statement, while the customer-harm cost of a false decline (lost lifetime value, reputational damage, the customer's switching cost to a competitor) is diffuse, delayed, and frequently absent from the objective the model is actually tuned against. An institution that reviews only $c_{FN}$-weighted accuracy metrics (fraud catch rate, basis-point loss reduction) is reviewing the model's performance, not its reward function; the reward function is the ratio $c_{FP}/c_{FN}$ implicit in the operating threshold, and that ratio is precisely the quantity Layer~1 review, per Section~\ref{sec:l1}, should require to be written down, approved, and version-controlled before deployment, exactly as a trading reward function's risk-aversion parameter would be.

\begin{table}[ht]
\centering\small
\caption{The fraud-decision-making cost matrix as the Layer~1 reward-function object}\label{tab:costmatrix}
\begin{tabular}{lcc}
\toprule
& Transaction legitimate & Transaction fraudulent\\
\midrule
Agent approves & 0 & $c_{FN}$ (fraud loss)\\
Agent declines & $c_{FP}$ (customer harm) & 0\\
\bottomrule
\end{tabular}
\end{table}

\paragraph{An unreviewed cost matrix is reward hacking, restated for fraud.} Section~\ref{sec:aisafety} frames reward hacking, following \citet{amodei2016} and \citet{skalse2022}, as an agent finding an unintended way to maximize a proxy objective that diverges from the designer's true intent. A fraud model tuned to minimize an implicit cost matrix that under-weights $c_{FP}$ relative to the institution's actual risk tolerance is not malfunctioning by any accuracy metric it is being evaluated against; it is doing exactly what its unreviewed objective specifies, declining aggressively because the training process never priced the customer-harm side of the matrix at its true cost. This is the fraud-domain instance of Section~\ref{sec:pdf}'s compliance-as-PDF problem: an institution's stated risk appetite (a target false-decline rate, a target customer-experience threshold) can exist as a policy document while the deployed model optimizes a materially different, unreviewed cost ratio, and nothing in the model's execution path enforces the policy document the moment the two diverge.

\paragraph{The fair-lending exposure specific to this domain.} Section~\ref{sec:l1} does not need to address disparate impact, because a trading reward function has no analog: a mispriced risk-aversion parameter produces poor risk-adjusted returns, not a protected-class harm. A fraud cost matrix does have this analog, and it is the central reason we propose cost-matrix review as a named, mandatory Layer~1 requirement for fraud pipelines rather than treating it as a special case of the base requirement. A cost matrix tuned purely to minimize $c_{FN}$ (fraud loss minimization, unconstrained) can learn to reduce false negatives by increasing false positives disproportionately among customer segments correlated with a protected characteristic (geography, name origin, spending pattern), precisely the mechanism formalized as the fraud-domain worked example in Section~\ref{sec:l2fraud}, where the drifted regime's decline signal begins reacting to a feature standing in for a protected characteristic. That worked example demonstrates detection of this drift after the fact via the integrated-covariance monitor, and exact attribution of the resulting regret to the affected segment; cost-matrix review under this extended Layer~1 is the corresponding pre-deployment control, intended to catch an unreviewed cost asymmetry before it produces the drift Section~\ref{sec:l2fraud}'s monitor is built to catch in production. The two controls are linked more tightly than by analogy: the monitor's regret is denominated in the reviewed cost matrix, so a mispriced $c_{FP}$ makes the monitor itself under-react to customer harm (Section~\ref{sec:l2fraud}).

\paragraph{Extended Layer 1 review requirement.} We propose that Layer~1 review for any fraud-decision-making agent requires, prior to deployment and at each retraining cycle:
\begin{enumerate}
\item An explicit, documented value for $c_{FP}$ and $c_{FN}$ (or the ratio between them), reviewed and approved with the same independence that Section~\ref{sec:l1} requires of a trading reward function, by a function distinct from the team that owns the model's accuracy metrics.
\item A disparate-impact analysis of the false-positive rate across protected-class-correlated segments, conducted against the reviewed cost matrix rather than the model's raw accuracy, since reviewing accuracy alone, as Section~\ref{sec:pdf} argues generally, evaluates compliance with a policy document rather than the policy actually executing in the production path.
\item Version control of the cost matrix itself, independent of model version control, so that a shift in $c_{FP}/c_{FN}$ introduced at a retraining cycle (Section~\ref{sec:satax}'s $M=2$--3 mutability rating for fraud agents, Table~\ref{tab:tax_sa}) is auditable as a policy change in its own right, distinguishable from a routine model refresh.
\end{enumerate}

\paragraph{Relation to the other extended layers.} Cost-matrix review is the pre-deployment complement to the in-production controls proposed elsewhere in this extension: it is what Layer~2's kill-switch (Section~\ref{sec:l2kill}) and drift monitor (Section~\ref{sec:l2fraud}) are checking a live model against, the validated cost ratio the model was reviewed and approved to implement. A kill-switch calibrated against a validated regime that itself embeds an unreviewed disparate-impact exposure will faithfully detect drift away from that regime while never flagging the exposure that was present in the regime from the outset, which is precisely why Section~\ref{sec:l1}'s review discipline must be applied to the cost matrix before deployment rather than inferred after the fact from the monitor's own calibration window.

\subsection{Layer 2: Engineering Layer (What You Enforce)}\label{sec:l2}
We require (a) policy stability monitoring, (b) compliance agents, and (c) kill-switch architecture. Policy stability monitoring can be implemented on the integrated-covariance identity of \citet{aldridge2026}. Let the deployed decision rule $\pi_t$ select, in period $t$,
\[
z_t=\pi_t(c)\in\arg\min_{z\in\mathcal Z}\; c^\top z+h_t(z),\qquad V_t(c):=\min_{z\in\mathcal Z}\; c^\top z+h_t(z),
\]
for a cost vector $c$, a feasible set $\mathcal Z$ that may be discrete (so binary approve/decline decisions are covered), and a term $h_t$ that collects everything else in the objective (risk penalties, other inputs). Let $\hat c_t$ be the reference cost on which the decision is evaluated (the forecast the decision was taken on, or the calibration-window mean), $c_t$ the realized cost, and $\Delta_t=c_t-\hat c_t$. The loss of acting on $\hat c_t$, $\ell_t:=c_t^\top\pi_t(\hat c_t)+h_t(\pi_t(\hat c_t))-V_t(c_t)\geq 0$, satisfies, exactly and scenario by scenario,
\begin{equation}\label{eq:ic}
\ell_t=-\int_0^1 \Delta_t^\top\bigl[\pi_t(\hat c_t+u\Delta_t)-\pi_t(\hat c_t)\bigr]\,du,
\qquad 0\leq \ell_t\leq C_t:=-\Delta_t^\top\bigl[\pi_t(c_t)-\pi_t(\hat c_t)\bigr].
\end{equation}
Cumulative regret over a monitoring horizon is therefore
\begin{equation}\label{eq:icT}
\Regret^{(T)}(\Pi)=\sum_{t=1}^T\ell_t=-\sum_{t=1}^T\int_0^1 \Delta_t^\top\bigl[\pi_t(\hat c_t+u\Delta_t)-\pi_t(\hat c_t)\bigr]\,du\;\leq\;\sum_{t=1}^T C_t .
\end{equation}
With $\hat c_t=\mu:=\E[c]$ and $c_u:=\mu+u(c-\mu)$, taking expectations gives the population form
\begin{equation}\label{eq:icpop}
\Regret=-\int_0^1\frac{\Cov\bigl(c_u,\pi(c_u)\bigr)}{u}\,du,\qquad 0\leq\Regret\leq-\Cov\bigl(c,\pi(c)\bigr),
\end{equation}
where $\Cov(c,\pi(c)):=\E[(c-\mu)^\top\pi(c)]$ is the scalar cost--decision covariance. Regret is the covariance of inputs and outputs, accumulated as uncertainty is switched on from zero ($u=0$) to its actual level ($u=1$). A cost-minimizing rule moves against its cost, so the covariance is nonpositive and enters with a minus sign. Four properties make \eqref{eq:ic}--\eqref{eq:icpop} a governance statistic:
\begin{enumerate}[label=(\roman*)]
\item \textbf{Exactness without regularity.} The identity needs only that decisions are produced by an optimizer; it requires no smoothness or continuity of $\pi_t$, so the same estimator applies to continuous position sizing and to a discontinuous approve/decline threshold. Because it holds for any reference $\hat c_t$, prediction bias is absorbed exactly: no separate baseline-drag term has to be assumed or held fixed.
\item \textbf{A conservative, passive screen.} The plain covariance is always an upper bound on regret. It can be computed from observed (cost, decision) pairs alone, for any vendor-supplied strategy, from the first day of deployment.
\item \textbf{A closed form for affine decision layers.} When $\pi(c)=Mc+m_0$, regret is exactly half the covariance, $-\tfrac12\tr(M\Sigma)$, with $\Sigma=\Cov(c)$ and no solves. This covers unconstrained and equality-constrained quadratic programs, including budget-constrained mean--variance portfolio construction \citep[Theorem~6 and Corollary~7]{aldridge2026}. For purely linear objectives an affine rule must be constant, so the half-covariance shortcut does not apply to threshold rules.
\item \textbf{Two diagnostics.} The ratio $\rho:=\Regret/(-\Cov)\in[0,1]$ equals $\tfrac12$ for affine rules; it rises above $\tfrac12$ when decisions respond early and saturate (for instance, when a hard limit binds) and falls below it when decisions respond only to large perturbations or are systematically, rather than randomly, wrong. The decision-stability margin $u^\star$, the largest fraction of the actual uncertainty under which the decision is on average unaffected, tightens the bound to $\Regret\leq(1-u^\star)(-\Cov)$.
\end{enumerate}

Table~\ref{tab:tiers} arranges the estimators by the access they require. None requires model weights or internal state. The passive tier is what an institution can compute for any vendor-supplied or vendor-embedded strategy (Section~\ref{sec:l25}) immediately; the replay tier requires sandboxed query access to the decision layer, re-running it on counterfactual cost inputs with all other inputs held fixed, which Layer~3 (Section~\ref{sec:l25}, item~6) makes a contractual requirement for vendor-supplied components. The identity presumes that decisions come from an optimizer of an objective linear in the cost with $h_t$ fixed over the evaluation window. A decision layer that is not one is itself detectable: along every replay path the covariance profile must be nondecreasing and bounded by $C_t$ (Algorithm~\ref{alg:monitor}, step~9), so a violation flags either an implementation error or a decision layer that is not optimizing any fixed objective of this form, which is itself a Layer~2 integrity alarm.

\begin{table}[ht]
\centering\small
\caption{Layer~2 regret estimators by required access, following \citet[Table~1]{aldridge2026}. $n$ is the number of observations in the evaluation window and $K$ the number of quadrature nodes. No tier requires model weights or internal state.}\label{tab:tiers}
\begin{tabularx}{\textwidth}{>{\raggedright\arraybackslash}p{3.6cm}>{\raggedright\arraybackslash}X>{\raggedright\arraybackslash}p{2.0cm}>{\raggedright\arraybackslash}X}
\toprule
Estimator & Access required & Decision-layer calls & Output\\
\midrule
Radius bound $D\sqrt{\tr\hat\Sigma}$ & Realized costs; diameter $D$ of the feasible set & 0 & Upper bound\\
Passive covariance $-\widehat{\Cov}(c,z)$ & Observed (cost, decision) pairs & 0 & Upper bound; exactly $2\times$ regret if the decision layer is affine\\
Closed form $-\tfrac12\tr(M\hat\Sigma)$ & Known affine decision layer (e.g., the institution's own budget-constrained mean--variance step) & 0 & Exact regret\\
Integrated covariance (Algorithm~\ref{alg:monitor}) & Replay (query) access to the decision layer & $nK+n+1$ & Exact regret, covariance bound, profile, $\hat\rho$, $\hat u^\star$, integrity check\\
\bottomrule
\end{tabularx}
\end{table}

Kill-switch triggers should be built on inner rather than declared LLM confidence, per \citet{chen2026}, since declared confidence is biased by the decoding process, while inner (pre-decoding) confidence is empirically informative about realized accuracy.

\subsubsection{Layer 2 Extended: Detecting Policy Drift via Integrated Covariance}\label{sec:l2drift}
The property Layer~2 requires is that an institution be able to monitor a vendor-supplied or otherwise opaque policy. To demonstrate that \eqref{eq:icT} is genuinely computable from observed data, at the passive tier from costs and decisions alone and at the replay tier with query access but no weight access, we construct a synthetic environment with a known, unmonitored change point and show that the statistic detects it.

\paragraph{Synthetic environment.} Over $T=500$ periods, we generate a realized cost shock $c_t\sim N(0,1)$ (implementation cost per unit of size) and other decision inputs $\eta_t\sim N(0,0.5^2)$. The sizing decision is
\begin{equation}\label{eq:tradingrule}
z_t=\pi_t(c_t)=\clip(\eta_t-\theta_t c_t,\,-L,\,L),\qquad L=1.5,
\end{equation}
which is the solution of $\min_{|z|\leq L}\,c_t z+(z-\eta_t)^2/(2\theta_t)$: a box-constrained quadratic program in which $L$ is the desk's rules-engine position limit (Section~\ref{sec:rules}) and $\theta_t$ is the policy's reactivity to cost. Before $t=300$, $\theta_t=\theta_0=0.2$: the validated policy trims size modestly when costs are high. From $t=300$ onward, $\theta_t=\theta_1=1.2$, a stylized form of reward hacking: the system has learned to condition its sizing on realized slippage far more aggressively than was ever validated. The change operates entirely inside the rules-engine limit, and it need not show up as a deterioration in P\&L, since on its own objective the reactive policy may well look better. It manifests only as a shift in the co-movement between costs and decisions, which is what \eqref{eq:ic} measures. Drift can also run the other way. In the RL trading agents of \citet{dou2025collusion}, the learned behavior is a collusive \emph{under}-reaction to information, which would lower the integrated covariance rather than raise it; an institution concerned with that failure mode should run the monitor with a two-sided trigger, $|\widehat\Regret_t-\mu_{\text{validated}}|>k\,\sigma_{\text{validated}}$.

\paragraph{What regret measures here.} For this rule, regret is the expected loss, in the policy's own objective units, of forcing it to act on the mean cost rather than the realized one: the economic stake the policy has placed on reacting to costs. In population, the validated policy's regret is $0.10$, which equals $\tfrac12\theta_0\Var(c)$ because the limit essentially never binds and the rule is affine ($\rho=0.50$). The drifted policy's regret is $0.53$, with covariance bound $0.90$ and $\rho=0.58$: the limit binds in 25\% of periods, the response saturates, and the covariance profile becomes concave. The half-covariance shortcut that is exact for affine layers would understate the drifted regret by 14\%, while the plain covariance remains a valid upper bound.

\paragraph{Estimators.} We compute two monitors on a trailing window of $W=20$ periods ending at $t$:
\begin{itemize}
\item \emph{Passive tier.} $-\widehat{\Cov}_t$, the sample covariance of the observed $(c_\tau,z_\tau)$, halved as the affine-rule estimate of regret. It needs no access beyond the observed series.
\item \emph{Replay tier} (Algorithm~\ref{alg:monitor}). The reference is the window mean $\bar c_t$; the decision layer is re-queried at $\bar c_t+u_k(c_\tau-\bar c_t)$ with the other inputs $\eta_\tau$ held fixed, at 16 Gauss--Legendre nodes $u_k$. This returns the regret $\widehat\Regret_t$, the covariance bound $\hat C_t$, and $\hat\rho_t$.
\end{itemize}
For each, the trigger level is $\tau=\mu_{\text{validated}}+k\,\sigma_{\text{validated}}$ with $k=4$, calibrated on $t\in[W,200)$; the remainder of the validated regime, $t\in[200,300)$, is held out to measure false alarms out of sample.

\begin{algorithm}[ht]
\caption{Layer~2 policy-stability monitor (integrated covariance)}\label{alg:monitor}
\begin{algorithmic}[1]
\State \textbf{Input:} observed costs $\{c_t\}$ and decisions $\{z_t\}$; reference costs $\{\hat c_t\}$ (default: trailing-window mean); optional replay oracle for $\pi_t$; window $W$; quadrature nodes $u_1<\dots<u_K$ with weights $w_k$; validated calibration window; $k$
\For{each period $t$, with window $\mathcal W_t=\{t-W+1,\dots,t\}$}
  \State $P_t\gets-\frac1W\sum_{\tau\in\mathcal W_t}(c_\tau-\bar c_t)^\top(z_\tau-\bar z_t)$ \Comment{passive tier: upper bound; $\tfrac12P_t$ exact if affine}
  \If{replay access}
    \State $\Delta_\tau\gets c_\tau-\hat c_\tau$,\; $\hat z_\tau\gets\pi_\tau(\hat c_\tau)$ for $\tau\in\mathcal W_t$
    \State $\hat\gamma_{t,k}\gets-\frac1W\sum_\tau\Delta_\tau^\top\bigl[\pi_\tau(\hat c_\tau+u_k\Delta_\tau)-\hat z_\tau\bigr]$, $k=1,\dots,K$ \Comment{covariance profile}
    \State $\hat C_t\gets-\frac1W\sum_\tau\Delta_\tau^\top\bigl[\pi_\tau(c_\tau)-\hat z_\tau\bigr]$ \Comment{covariance bound}
    \State $\widehat\Regret_t\gets\sum_k w_k\hat\gamma_{t,k}$;\quad $\hat\rho_t\gets\widehat\Regret_t/\hat C_t$;\quad $\hat u^\star_t\gets\max\bigl(\{0\}\cup\{u_k:\hat\gamma_{t,k}=0\}\bigr)$
    \State \textbf{check} $0\leq\hat\gamma_{t,1}\leq\dots\leq\hat\gamma_{t,K}\leq\hat C_t$; otherwise raise an integrity alarm
  \EndIf
\EndFor
\State Calibrate $\tau\gets\mu_{\text{validated}}+k\,\sigma_{\text{validated}}$ for the chosen statistic ($\mu_{\text{validated}}-k\,\sigma_{\text{validated}}$ for $\hat\rho_t$)
\For{each new period $t$}
  \If{the statistic crosses $\tau$}
    \State raise drift alarm; escalate to Layer~1 review
  \EndIf
\EndFor
\end{algorithmic}
\end{algorithm}

\begin{figure}[ht]
\centering
\includegraphics[width=0.95\textwidth]{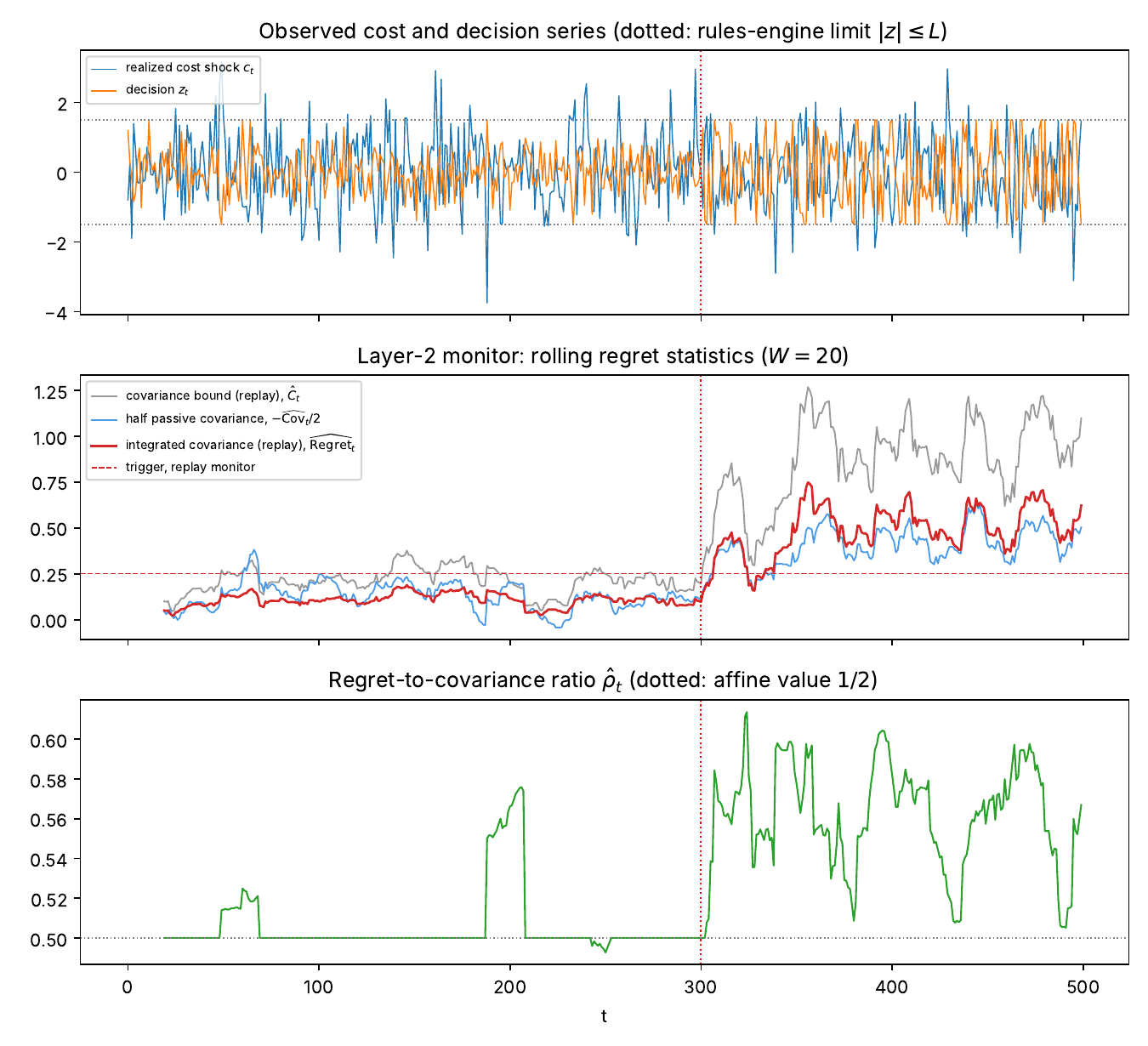}
\caption{Top: observed cost shock $c_t$ and decision $z_t$; the true (monitor-unobserved) drift onset at $t=300$ is marked. Middle: rolling regret statistics computed from the observed series (passive) and from replay of the decision layer (integrated covariance and covariance bound); the replay statistic crosses its trigger 7 periods after onset. Bottom: the regret-to-covariance ratio, equal to the affine value $\tfrac12$ except where the rules-engine limit binds.}\label{fig:trading}
\end{figure}

\paragraph{Result.} Both statistics remain low throughout the validated regime and rise after the change point at $t=300$ (Figure~\ref{fig:trading}). Table~\ref{tab:trading} reports the summary statistics. The replay monitor raises its first alarm 7 periods after true drift onset; the passive monitor, 55 periods after. The replay tier is faster because it compares each decision with its own counterfactual, so the noise contributed by the other inputs $\eta_t$ cancels, whereas the passive covariance must average that noise out. In both cases drift onset is detected from the observed cost/decision series and, for the replay tier, from queries to the decision layer, with no access to the policy's internal parameters. The ratio $\hat\rho_t$ sits at $\tfrac12$ throughout the validated regime except in the few windows in which the position limit binds, and rises toward $0.58$ after drift, as the limit binds more often: without any access to the rules engine, the monitor reports how much of the policy's behavior the hard limit is now absorbing. As a check on the identity, the quadrature value of \eqref{eq:ic} matches the directly computed loss to within $7\times10^{-4}$ relative error in every window, and the profile is monotone in every window, as the identity requires.

\begin{table}[ht]
\centering\small
\caption{Integrated-covariance monitor, trading domain: summary statistics ($W=20$, $k=4$, seed 2026). Regime means use windows lying entirely in each regime.}\label{tab:trading}
\begin{tabular}{lcc}
\toprule
Quantity & Replay tier: $\widehat\Regret_t$ & Passive tier: $-\widehat{\Cov}_t/2$\\
\midrule
Mean statistic, validated regime ($t<300$) & 0.107 & 0.128\\
Mean statistic, drifted regime ($t\geq320$) & 0.502 & 0.413\\
Ratio (drifted / validated) & $4.7\times$ & $3.2\times$\\
Trigger level ($\mu_{\text{validated}}+4\sigma_{\text{validated}}$) & 0.251 & 0.445\\
True drift onset & $t=300$ & $t=300$\\
First alarm at or after onset & $t=307$ & $t=355$\\
Detection latency & 7 periods & 55 periods\\
False-alarm rate, hold-out $t\in[200,300)$ & 0.00 & 0.00\\
Mean $\hat\rho_t$, validated / drifted & 0.51 / 0.56 & ---\\
Population regret, validated / drifted & \multicolumn{2}{c}{0.10 / 0.53 \quad (covariance bound 0.20 / 0.90)}\\
\bottomrule
\end{tabular}
\end{table}

This synthetic example is a calibrated illustration, not a claim to reproduce any specific vendor model's internal behavior. The parameters ($W$, $\theta_0$, $\theta_1$, $L$, $k$) are stated explicitly and can be re-swept against real cost/decision data by an adopting institution. Appendix~\ref{app:robust} reports that this single-seed result is representative: across a grid of $(W,\theta_1,k)$ and 30 independent seeds per configuration, the replay monitor detects every drift, with a mean latency of 4--11 periods, and the passive monitor is roughly three times slower.

\subsubsection{Layer 2 Extended: Inner-Confidence Kill-Switches for Fraud-decision-making Agents}\label{sec:l2kill}
Section~\ref{sec:l2} requires that kill-switch triggers be based on an agent's \emph{inner} confidence rather than its \emph{declared} confidence, following \citet{chen2026}, since declared confidence is biased by the decoding process, while inner (pre-decoding) confidence is empirically informative about realized accuracy. This distinction is not merely useful for a fraud-decision-making agent; it is load-bearing in a way that it is not for a trading strategy. A trading counterparty does not observe and does not optimize against an agent's live confidence score. A fraudster does. Any exposed fraud-probability score, whether returned directly to a merchant API or inferable from repeated decline/approve outcomes, is a signal that an adversarial actor can probe and route around. A kill-switch keyed to the declared score is therefore not merely biased in the way \citet{chen2026} documents for LLM decoding. Instead, it is a control surface that an adversary has a direct financial incentive to learn and defeat. This is a fraud-specific sharpening of the adversarial-input-distribution problem posed generally in Section~\ref{sec:adversarial}.

\paragraph{Design.} Define $\hat p_t\in[0,1]$ as the agent's declared fraud-probability output for transaction $t$, and $z_t$ as an inner-confidence signal computed upstream of the decision layer, for instance, the entropy of the model's pre-decision logits or the distance of the transaction's learned embedding from the centroid of the validated training distribution. Whereas $\hat p_t$ is the quantity a fraud ring can iteratively probe by varying transaction structuring, $z_t$ is not directly observable from decision outcomes alone, since it is never returned to the caller and does not correspond one-to-one with the approve/decline boundary that an adversary can map empirically.

The Layer~2 trigger condition becomes:
\begin{equation}\label{eq:kill}
\text{Kill-switch fires at } t \iff z_t<\tau_z,
\end{equation}
where $\tau_z$ is calibrated, exactly as in Algorithm~\ref{alg:monitor}'s $\tau\leftarrow\mu_{\text{validated}}\pm k\,\sigma_{\text{validated}}$, against a known-validated calibration window of transactions, rather than against a fraud-loss target, since the latter is precisely the quantity under adversarial pressure to be minimized. Unlike the integrated-covariance monitor of Section~\ref{sec:l2drift}, which needs only costs and decisions, $z_t$ is computed from data internal to the model's forward pass; like it, $z_t$ is not the quantity the model was trained to expose or optimize. A complementary, model-external margin is the per-transaction decision-stability margin of Section~\ref{sec:l2fraud}, which measures how far a decision sits from the decline boundary relative to the size of its realized error; the transactions with the smallest margins are exactly those an adversary probing the boundary is searching for.

\paragraph{Adapted monitor.} Algorithm~\ref{alg:kill} restates the Layer~2 monitor for the fraud setting. The essential change is not mechanical. In Algorithm~\ref{alg:kill}, the calibration window and trigger threshold must be re-derived on a substantially shorter cadence than in the trading case, since fraud rings adapt their structuring strategies on the order of days to weeks rather than the multi-month drift horizon documented for MSRR in Section~\ref{sec:msrrdesc}. A Layer~2 monitor, calibrated once at deployment and left static, represents exactly the static-validation failure mode Section~\ref{sec:static} identifies as the root cause of the governance gap. This is a materially weaker control here than in the trading case because the input distribution is actively and intentionally adversarial rather than merely nonstationary.

\begin{algorithm}[ht]
\caption{Layer~2 fraud kill-switch monitor (inner-confidence)}\label{alg:kill}
\begin{algorithmic}[1]
\State \textbf{Input:} inner-confidence series $\{z_t\}$ (pre-decoding, not exposed to callers), declared score series $\{\hat p_t\}$, validated-regime calibration window, recalibration interval $\Delta$
\State Compute trigger level $\tau_z\gets\mu_{\text{validated}}-k\cdot\sigma_{\text{validated}}$ from a known-validated calibration window
\For{each new transaction $t$}
  \If{$z_t<\tau_z$}
    \State route decision to human review or default to safe action (hold, not auto-decline)
    \State escalate to Layer~1 review
  \EndIf
  \If{$t \bmod \Delta=0$}
    \State recalibrate $\tau_z$ against most recent validated window
    \State flag for audit if $\tau_z$ has drifted materially from prior calibration
  \EndIf
\EndFor
\end{algorithmic}
\end{algorithm}

\paragraph{Why the fallback action differs from the trading case.} In Section~\ref{sec:l2}'s trading context, a kill-switch firing halts new position-taking, a symmetric, low-cost default. In the fraud context, the two candidate default actions, auto-decline or auto-approve, carry asymmetric and non-substitutable costs: routing every low-inner-confidence transaction to automatic decline reintroduces the false-decline harm that the agent was deployed to reduce, while defaulting to approval defeats the control's purpose. We therefore specify the default action as routing to human review rather than either automated extreme, which is the fraud-domain analog of Section~\ref{sec:sal2}'s broader principle that a governed system needs a mechanism that can force de-risking before the system's own dynamics do so involuntarily, at a worse price. Here, the worse price is a customer harmed by an automatic decline rather than a book unwound at a discount.

\paragraph{Relation to Layer 3.} Where the fraud-decision-making pipeline itself is compositional, with an identity-verification vendor, a device-fingerprinting vendor, and an in-house or vendor-supplied scoring model, in the sense of Section~\ref{sec:l25fraud}, the inner-confidence signal $z_t$ should be computed and monitored separately at each stage rather than only at the final decision output, since a vendor-side update to an upstream component (Section~\ref{sec:l25fraud}'s vendor model-version attestation requirement) can degrade the reliability of $z_t$ itself without triggering a decision-level alarm, exactly the composition-level blind spot described in Section~\ref{sec:composition}.

\subsubsection{Layer 2 Extended: Detecting Fraud-Model Policy Drift via Integrated Covariance}\label{sec:l2fraud}
Section~\ref{sec:l2drift} demonstrates that \eqref{eq:icT} is genuinely computable from observed data in a trading context. The identical identity transfers to a fraud-decision-making agent with no change to the estimator, only to the interpretation of $c_t$ and $\pi_t$, and with one change of emphasis: an approve/decline rule is a threshold, so the decision layer is piecewise constant and the integrated covariance, not the half-covariance shortcut, is the exact object \citep[Proposition~8]{aldridge2026}. We construct a synthetic environment with a known, unmonitored change point representing a specific and realistic fraud-governance failure: a model that silently begins over-indexing on a spurious, protected-class-correlated feature after a retrain. We show that the monitor detects it without access to the model's internal parameters and without knowing in advance which feature to look at.

\paragraph{Synthetic environment.} Over $T=500$ periods (transaction batches) of $n_b=1{,}000$ transactions, each transaction has a legitimate risk feature $x\sim N(0,1)$, a binary segment indicator $g$ (share 0.3) that carries no fraud signal but stands in for a protected characteristic (e.g., a merchant category code or a geographic cluster), and a fraud label $y\sim\text{Bernoulli}(\sigma(-3.8+1.2x))$, a base rate of about 4\%. The model's predicted fraud probability $\hat p$ is the true logit plus $N(0,0.5^2)$ estimation noise. The decision layer is the cost matrix of Table~\ref{tab:costmatrix} with $c_{FN}=1$ and $c_{FP}=0.1$. Per transaction, the realized net cost of declining is $c=c_{FP}(1-y)-c_{FN}\,y$, its prediction is $\hat c=c_{FP}-(c_{FP}+c_{FN})\hat p$, and the agent declines if and only if $\hat c<0$, that is, $\hat p>c_{FP}/(c_{FP}+c_{FN})\approx0.091$: a linear program on $[0,1]$ per transaction, with a threshold decision rule. From $t=300$ onward, after a retrain, the model's logit acquires a spurious term $\theta(g-0.3)$ with $\theta=1.2$:
\begin{equation}\label{eq:fraudrule}
\operatorname{logit}\hat p=-3.8+1.2x+\varepsilon+\theta(g-0.3)+\delta(\theta),\qquad \varepsilon\sim N(0,0.5^2),
\end{equation}
where, as is standard operational practice after a retrain, the decision threshold is recalibrated to hold the aggregate decline rate at its validated level (a global logit offset $\delta(1.2)=-0.14$). This is a stylized form of the fraud-domain reward-hacking failure described in Section~\ref{sec:l1fraud}: the model has learned to condition its declines on a feature that proxies for a protected characteristic, in a way that was never validated.

\paragraph{Why aggregate checks miss it.} The aggregate decline rate is unchanged by construction (12.5\% in both regimes, in population), and fraud loss per transaction moves from 0.023 to 0.024, well inside batch-to-batch noise. Meanwhile, the false-decline rate among legitimate customers in segment $g=1$ rises from 11\% to 25\%, while falling to 5\% in segment $g=0$. Overall fraud-catch performance is essentially stable while the basis for individual declines shifts.

\paragraph{What the identity computes for a threshold rule.} For a threshold rule the integral in \eqref{eq:ic} evaluates in closed form: the loss equals the misclassification cost priced by the reviewed cost matrix ($c_{FP}$ for a false decline, $c_{FN}$ for a missed fraud, zero otherwise), and the covariance term is $C=|\Delta|$ for a misclassified transaction. Per transaction, $\ell/C=1-u_{\text{flip}}$, where $u_{\text{flip}}$ is the fraction of the realized surprise $\Delta$ that the prediction absorbs before the decision flips: a transaction-level decision-stability margin. The batch ratio $\hat\rho=\sum\ell/\sum C$ therefore measures how far, on average, wrong decisions sat from the decision boundary relative to the size of the error. It is high when errors are noise around the boundary, and it falls when errors become systematic, that is, when declines become confidently wrong. \citet{aldridge2026} report the same signature for a misspecified predictor in the shortest-path benchmark, where biased errors raise the covariance term without a commensurate rise in loss.

\paragraph{Monitor.} We pool the trailing $W=20$ batches and compute, per transaction, the regret $\widehat\Regret_t$ (Algorithm~\ref{alg:monitor} with $\hat c$ equal to the model's prediction), the covariance term $\hat C_t$, and $\hat\rho_t$. The primary statistic is $\hat\rho_t$, with a lower trigger $\mu_{\text{validated}}-k\,\sigma_{\text{validated}}$, $k=3$, calibrated on $t\in[W,200)$ and with false alarms measured on $t\in[200,300)$. For comparison we run the same trigger rule on total regret, on the covariance term, and on the two aggregate metrics an accuracy-focused review would track. Realized labels are needed, so in production the monitor runs on the chargeback lag (weeks); the inner-confidence kill-switch of Section~\ref{sec:l2kill} covers that interval.

\begin{figure}[ht]
\centering
\includegraphics[width=0.95\textwidth]{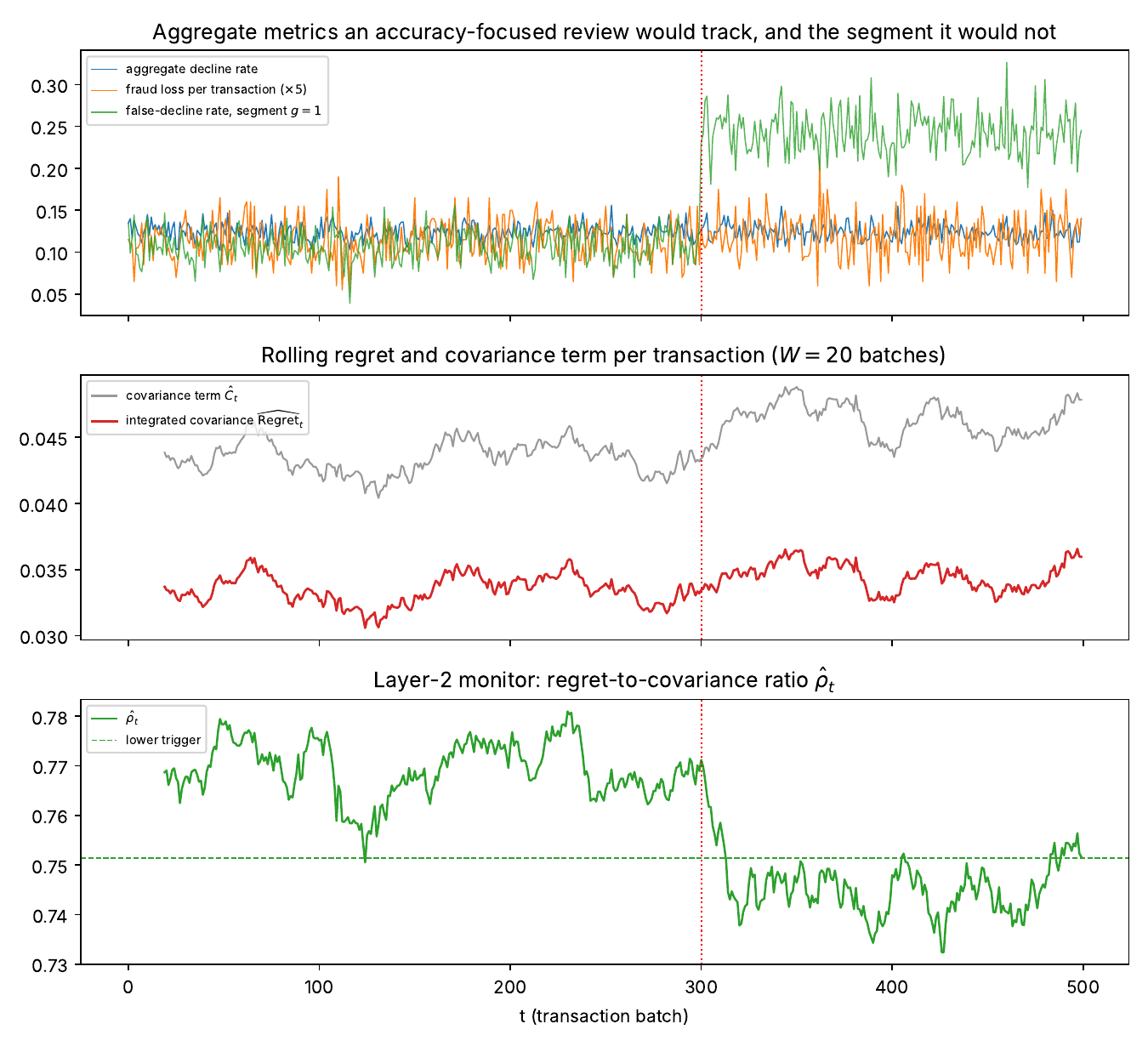}
\caption{Top: per-batch aggregate decline rate and fraud loss, which a fraud-loss-focused review would track, and the false-decline rate in segment $g=1$, which it would not; the true (monitor-unobserved) drift onset at $t=300$ is marked. Middle: rolling integrated-covariance regret and covariance term per transaction. Bottom: the regret-to-covariance ratio $\hat\rho_t$, computed from predictions, decisions and realized labels alone, crosses its lower trigger 13 batches after onset.}\label{fig:fraud}
\end{figure}

\paragraph{Result.} Table~\ref{tab:fraud} and Figure~\ref{fig:fraud} report the results. The ratio $\hat\rho_t$ falls from 0.769 to 0.745 and crosses its trigger 13 batches after drift onset. The covariance term rises 7\% and crosses its trigger after 19 batches. Total regret rises only 3\% and never crosses its trigger in the 200 post-drift batches; neither do the aggregate decline rate and fraud loss. Total regret barely moves for two reasons: the drift redistributes false declines across segments rather than adding them, and the reviewed cost matrix prices a false decline at one-tenth of a fraud loss, so the irreducible fraud-side loss dominates the total. What the drift changes is the \emph{character} of the errors, which is what $\rho$ measures. This is also a direct quantitative link to Layer~1: the regret monitor is denominated in the reviewed cost matrix and inherits any mispricing of $c_{FP}$, which is why cost-matrix review (Section~\ref{sec:l1fraud}) has to come first. As a check, a 400-node midpoint evaluation of \eqref{eq:ic} matches the closed-form loss to within $6\times10^{-6}$ relative error, and $0\leq\ell\leq C$ holds for every transaction.

\begin{table}[ht]
\centering\small
\caption{Integrated-covariance monitor, fraud domain: summary statistics ($W=20$ batches of 1,000 transactions, $k=3$, seed 2026). Values are per transaction; regime means use windows lying entirely in each regime. True drift onset $t=300$; no statistic alarmed in the hold-out window $t\in[200,300)$.}\label{tab:fraud}
\begin{tabular}{lcccc}
\toprule
Statistic & Validated mean & Drifted mean & Trigger & Detection latency\\
\midrule
$\hat\rho_t$ (primary; lower trigger) & 0.769 & 0.745 & 0.751 & 13 batches\\
Covariance term $\hat C_t$ & 0.0435 & 0.0464 & 0.0471 & 19 batches\\
Total regret $\widehat\Regret_t$ & 0.0334 & 0.0346 & 0.0368 & no alarm\\
Aggregate decline rate & 0.124 & 0.124 & 0.130 & no alarm\\
Fraud loss ($c_{FN}\times$ missed fraud) & 0.0227 & 0.0238 & 0.0258 & no alarm\\
\bottomrule
\end{tabular}
\end{table}

\paragraph{Exact attribution to segments.} Because \eqref{eq:ic} holds scenario by scenario, regret, the covariance term and $\rho$ can be averaged over any partition of transactions, and the averages are exact. The monitor needed no segment information to fire; once it has, the Layer~1 reviewer can scan candidate segments, including ones the original fair-lending review did not know to look for. Table~\ref{tab:segments} shows the population values. The disparity is visible in the segment-level ratio even more sharply than in segment regret: in the disfavored segment $\rho$ falls from 0.77 to 0.58, and the average misclassified transaction now needs 39\% of its realized surprise, rather than 30\%, before its decision would flip. This is the disparate-impact analysis of Section~\ref{sec:l1fraud}, item~2, produced as a by-product of the Layer~2 monitor and denominated in the reviewed cost matrix.

\begin{table}[ht]
\centering\small
\caption{Segment attribution of integrated-covariance regret, fraud domain (population values, $4\times10^6$ transactions per regime, $\theta=1.2$). False-decline rate is among legitimate transactions; regret is per transaction in the segment.}\label{tab:segments}
\begin{tabular}{lcccc}
\toprule
& \multicolumn{2}{c}{Segment $g=1$ (proxy)} & \multicolumn{2}{c}{Segment $g=0$}\\
\cmidrule(lr){2-3}\cmidrule(lr){4-5}
& Validated & Drifted & Validated & Drifted\\
\midrule
False-decline rate & 11.2\% & 25.4\% & 11.2\% & 5.3\%\\
Regret & 0.0334 & 0.0381 & 0.0334 & 0.0337\\
\quad of which false-decline (customer-harm) regret & 0.0108 & 0.0244 & 0.0107 & 0.0051\\
$\rho$ & 0.77 & 0.58 & 0.77 & 0.87\\
Mean flip point $u_{\text{flip}}$, misclassified transactions & 0.30 & 0.39 & 0.30 & 0.23\\
\bottomrule
\end{tabular}
\end{table}

We note two fraud-domain-specific departures from the trading example. First, the trigger threshold here uses $k=3$ rather than $k=4$: fraud drift of this kind carries a direct compliance exposure (disparate-impact liability under fair-lending regulation) that a modest increase in the false-alarm rate is a reasonable price to avoid, whereas the trading example's $k=4$ reflects a context where a false alarm merely triggers a costly but non-regulatory review. Appendix~\ref{app:robust} makes the trade-off concrete: at $W=20$, raising $k$ from 3 to 4 doubles the mean detection latency (from 18 to 35 batches) and raising it to 5 misses one drift in five, while the out-of-sample false-alarm rate at $k=3$ stays below 0.5\%. Second, and consistent with the fraud-domain guidance of Section~\ref{sec:l2kill}, the calibration window should be re-swept substantially more frequently than in the trading case, since fraud rings adapt their structuring behavior over a horizon of days to weeks rather than the multi-month horizon over which a trading policy's inputs typically evolve.

This synthetic example is a calibrated illustration, not a claim to reproduce any specific institution's fraud model or feature set; the parameters ($W$, $\theta$, $k$, the cost matrix and the batch size) are stated explicitly and can be re-swept against real prediction, decision and chargeback data by an adopting institution, as in Section~\ref{sec:l2drift}. Appendix~\ref{app:robust} reports the same sensitivity analysis for this domain. Unlike the trading monitor, the fraud monitor's power is limited for weak drift: at $\theta=0.8$ it detects 83\% of drifts at $k=3$ and fewer at higher $k$.

\subsection{Layer 2, Extended: Bounded Action Authority for Customer-Facing Agents}\label{sec:l2bounded}
Section~\ref{sec:sal2} argues that Layer~2's underlying design principle, restated at the right level of abstraction, is that a governed system needs a mechanism that can force de-risking before the system's own dynamics do so involuntarily, at a worse price. For a retrained trading policy, that mechanism is the inner-confidence kill-switch of Section~\ref{sec:l2}. For a customer-facing conversational or action-taking agent (refund authorization, credit-limit adjustment, dispute resolution, account modification, service credits), the analogous mechanism is not a confidence threshold on the agent's output, but a hard ceiling on the scope of action the agent is authorized to execute unilaterally. We propose this as an explicit fourth Layer~2 control, \emph{bounded action authority}, alongside policy stability monitoring, compliance agents, and kill-switch architecture (Section~\ref{sec:l2}).

\paragraph{Why this control is distinct from the kill-switch.} The kill-switch of Section~\ref{sec:l2} is triggered by a signal about the model's internal state, inner confidence falling below a calibrated threshold. Bounded action authority requires no such signal and fires on no drift or confidence condition at all; it is a static, pre-committed ceiling on the consequence of any single action the agent is permitted to take without human sign-off, independent of how confident the agent is in that action. This distinction matters because a customer-facing agent can be highly confident and still be wrong in a way that a confidence-based trigger will never catch, for instance, an agent that confidently interprets an ambiguous customer request as authorization for an action the customer did not intend. A confidence threshold governs how sure the model is; an action ceiling governs how much damage a single wrong action can do, and the two must be implemented as separate controls because neither substitutes for the other.

\paragraph{Design.} For a customer-facing agent $s$, define an action $a$ with an associated consequence magnitude $v(a)$ (a refund amount, a credit-limit delta, a dispute-resolution award) expressed in a common unit, typically currency, though non-monetary actions such as account suspension or data-sharing changes should be scored on an institution-defined severity scale and included in the same ceiling framework. The control is:
\begin{equation}\label{eq:ceiling}
\text{Agent may execute } a \text{ unilaterally} \iff v(a)\leq\kappa(s),
\end{equation}
where $\kappa(s)$ is a per-agent, per-context ceiling set and reviewed independently of the team that owns the agent's development, the same independence that Section~\ref{sec:l1} requires of reward-function review. Actions with $v(a)>\kappa(s)$ are not blocked outright; they are routed to a human approver, mirroring the default-to-human-review design already proposed for the fraud kill-switch in Section~\ref{sec:l2kill}, rather than defaulting to either automatic denial or automatic execution, since both extremes reintroduce a version of the harm the control exists to prevent.

Consistent with the ARS taxonomy of Section~\ref{sec:taxonomy}, $\kappa(s)$ should scale inversely with the agent's own $(A,R)$ score: an agent scored $A=3$ (fully autonomous execution, no human in the loop at the point of action) and $R=1$--2 (the customer-facing agent row of Table~\ref{tab:tax_sa}) warrants a materially lower ceiling than one that only drafts an action for a human to confirm, since autonomy rises to $A=3$ the moment an agent's output is wired directly into execution (Section~\ref{sec:satax}), and a low-$R$ action is one for which a human is unwinding a harm rather than approving an action pre-emptively.

\paragraph{Aggregate ceiling and Layer 4 interaction.} A per-action ceiling alone is insufficient to bound the agent's aggregate exposure since an agent could execute many actions individually below $\kappa(s)$ that collectively produce a blast radius no single action would. We therefore add a rolling aggregate ceiling,
\begin{equation}\label{eq:aggceiling}
\sum_{a\in\mathcal A[t-\Delta,t]}v(a)\leq K(s),
\end{equation}
over a defined window $\Delta$, with a breach triggering the same kill-switch escalation path as Section~\ref{sec:l2}, halting further unilateral action pending a Layer~1 review. This is the direct fraud-domain-and-customer-service analog of Section~\ref{sec:sal2}'s leverage-tiered de-risking schedule: a per-position limit did not save Situational Awareness from an aggregate, firm-wide blast radius (Section~\ref{sec:satax}, $B=3$), and a per-action ceiling alone would fail a customer-facing agent for the identical structural reason. Where multiple customer-facing agents draw on a shared foundation model vendor or a shared action-execution layer, this aggregate ceiling should additionally be monitored at the portfolio level across agents, not merely per-agent, since correlated errors across agents built on the same underlying model are a Layer~4 crowding exposure (Section~\ref{sec:l3}) rather than an isolated Layer~2 one: a vendor-side update that shifts one agent's behavior plausibly shifts every agent built on the same foundation model simultaneously.

\paragraph{Relation to the reward function (Layer 1).} Section~\ref{sec:l1} requires the reward function to receive policy-document-grade review before deployment. For a customer-facing agent, $\kappa(s)$ and $K(s)$ are not incidental engineering parameters but are themselves expressions of institutional risk tolerance and should be reviewed and version-controlled alongside the reward function itself, with the same independence requirement: the team that owns the agent's conversational quality metrics should not be the sole owner of the ceiling that bounds its financial authority, for the same reason that Section~\ref{sec:sal1} argues that leverage sign-off should be separated from thesis sign-off.

\paragraph{Case grounding.} This control is not a hypothetical extension. Publicized incidents in which a customer-facing chatbot committed a firm to an action or representation outside its intended scope, and the firm was subsequently held to that commitment, are direct evidence of the underlying failure mode: an agent with unbounded action authority (or, more precisely, no explicit, reviewed ceiling at all) operating exactly as designed at the component level while producing a firm-level consequence that no single-turn audit of the conversation would have flagged. This is structurally the same pattern Section~\ref{sec:msrr} documents for MSRR (component-level correctness, composition-level failure) and Section~\ref{sec:sa} documents for Situational Awareness (no single decision was unreasonable in isolation; the aggregate exposure was), extended to the customer-facing setting, and it is the reason we propose bounded action authority as a named, mandatory Layer~2 control rather than leaving it implicit in the general kill-switch requirement of Section~\ref{sec:l2}.

\subsection{Layer 3: Interface and Composition Validation}\label{sec:l25}
Compositional risk does not fit cleanly into Layers 1, 2, or 4, each defined at the level of a single agent. Layer~3 takes the pipeline as its unit of governance, with six requirements: (1) pipeline-level adversarial red-teaming at the entry point, evaluated at the exit point; (2) interface contracts with explicit exclusion lists; (3) joint-output sampling in production, scaling with the pipeline's ARS; (4) a named accountable owner for the composition, distinct from component owners; (5) a vendor model-version attestation, documenting the vendor's model identifier and version for every production inference, so that changes can be traced to vendor updates, if any occurred; and (6) replay access to every vendor-supplied decision stage, so that the Layer~2 drift monitor can run at its exact tier.
\begin{enumerate}[label=(\arabic*)]
\item \textbf{Pipeline-level adversarial red-teaming,} evaluated at the exit point (the approve/decline/hold decision), in addition to each vendor's individual API boundary.
\item \textbf{Interface contracts with explicit exclusion lists,} specifying, for every third-party call in the pipeline and particularly any call routed through a foundation-model vendor, precisely which data fields are contractually excluded from that call: primary account numbers, full government ID numbers, and biometric templates should appear on this exclusion list by default. All exceptions should be individually justified and logged with the fields specified in advance rather than left generic.
\item \textbf{Joint-output sampling in production,} scaling with the pipeline's ARS score (Section~\ref{sec:taxonomy}), as in the base requirement, applied to the joint distribution of (identity-verification result, device-fingerprint result, fraud score, final decision) or a trading pipeline's joint return distribution.
\item \textbf{A named accountable owner for the composition,} distinct from each component vendor's relationship owner.
\item \textbf{Vendor model-version attestation, extended to regulated-data-scope attestation.} Any agentic stage of the pipeline that depends on a third-party foundation model must log the vendor's model identifier and version for every production inference so that a performance discontinuity can be traced to a vendor-side update rather than misattributed to a change in fraud patterns. For fraud pipelines specifically, this attestation must additionally record, for every such inference, whether any field on the exclusion list in item~(2) was present in the data passed to that vendor and must be auditable independently of the vendor's own compliance representations, since a vendor's model-card-style attestation of what data it claims not to retain is necessary but not sufficient due diligence: it establishes what the vendor intends, not what the deployed pipeline actually sent.
\item \textbf{Replay access for drift monitoring.} For any vendor-supplied stage that maps cost or risk inputs to decisions, the interface contract should guarantee sandboxed query access: the ability to re-run the stage, at the logged model version, on counterfactual cost inputs with all other inputs held fixed. This is the access the exact tier of the Layer~2 monitor needs (Table~\ref{tab:tiers}); it is far less than weight access and does not expose the vendor's intellectual property. Without it, the institution is limited to the passive covariance bound, which in Section~\ref{sec:l2drift} detected the same drift 55 rather than 7 periods after onset, and about three times more slowly on average across the robustness grid of Appendix~\ref{app:robust}. Replay also makes the monitor's integrity check (Algorithm~\ref{alg:monitor}, step~9) available, which tests whether the vendor's decision stage behaves as a consistent optimizer at all.
\end{enumerate}

\subsubsection{Layer 3: a concrete instance of multi-vendor compositional risk}\label{sec:l25msrr}
The MSRR pipeline is structurally a multi-module composition that motivates Layer~3: a data-retrieval/ingestion stage (news feed), an embedding stage (a third-party LLM: the paper itself documents results across BERT, GPT-2, Mistral-7B, and the Llama3 family at 8B/70B/405B parameters), a residualization stage (regression against stock characteristics), and a portfolio-construction stage (the MSRR optimization). \citet{didisheim2026} treat the choice of embedding model as a robustness check; from a governance perspective, it is the clearest illustration available of why Section~\ref{sec:composition}'s component-level audit is structurally insufficient. Each stage, audited in isolation, is unremarkable: the embedding model is a standard third-party encoder used as advertised, the residualization is an ordinary panel regression, and the portfolio optimization is a textbook mean--variance estimator. The composed pipeline, however, produces a return stream whose magnitude and behavior depend materially and nonlinearly on a component that the deploying institution typically does not control and frequently cannot inspect: the internal weights of the vendor's embedding model. A vendor-side model upgrade, the kind of change a deploying institution would not necessarily be informed of, let alone approve, can alter the joint output distribution of the pipeline without tripping any single-component drift monitor, exactly the failure mode Layer~3's joint output-sampling requirement (Section~\ref{sec:l25}, item~3) is designed to catch. The fifth Layer~3 requirement, motivated directly by this case, is vendor model-version attestation, under which any agentic pipeline that depends on a third-party foundation model must log the vendor's model identifier and version for every production inference, so that a performance discontinuity can be traced to a vendor-side update rather than misattributed to market regime change.

The composition also shows where the Layer~2 monitor can be anchored without any access to the vendor. The portfolio-construction stage belongs to the deploying institution, and the embedding stage only supplies its cost input (predicted returns). The scenario-wise identity \eqref{eq:ic} therefore applies with $\hat c_t$ equal to the pipeline's predicted cost vector, $c_t$ the realized one, and $\pi$ the institution's own optimizer, which it can always replay. To the extent that the construction step is an unconstrained or budget-constrained mean--variance optimizer, its decision rule is affine, $\pi(c)=Mc+m_0$, and the per-period regret of the vendor-driven signal is available in closed form, $\ell_t=-\tfrac12\Delta_t^\top M\Delta_t$, from the institution's own risk model and no solves \citep[Theorem~6 and Corollary~7]{aldridge2026}. A vendor-side embedding update that degrades or biases the predictions shows up directly in $\ell_t$, and a ratio $\hat\rho_t$ that departs from $\tfrac12$ indicates that inequality constraints (no-short or position limits) have started to bind. The vendor's weights never enter.

A second, related observation concerns lookahead bias. \citet{didisheim2026} show, following \citet{he2025}, that point-in-time and full-foresight versions of a chronologically consistent LLM produce statistically indistinguishable strategy performance, while replacing a small academic model with an industrial-scale model roughly doubles the Sharpe ratio. This is direct evidence for a claim Layer~1 and Layer~2 make on first principles: an agentic financial system's behavior is not fully characterized by testing its training data provenance alone; two models with identical training data vintage but different scales produce materially different live performance, so model-card-style attestation of training data cutoff, while necessary, is not sufficient Layer~2 due diligence for a vendor-supplied component.

\subsubsection{Vendor Attestation and Interface Contracts for Fraud-decision-making Pipelines}\label{sec:l25fraud}
The MSRR pipeline is structurally a multi-module composition, where each stage audited in isolation is unremarkable. However, the composed pipeline produces behavior that no component-level audit would surface. A production payments fraud-decision-making pipeline exhibits the identical structural property, chained through a data-ingestion stage (transaction and device telemetry), an identity-verification stage (a third-party KYC/identity vendor), a device-fingerprinting or behavioral-biometrics stage (typically a separate vendor), a sanctions/watchlist screening stage, a scoring stage (an in-house or vendor-supplied fraud model, increasingly LLM-assisted for narrative risk summarization or agentic case triage), and a decision-making stage that resolves these signals into an approve/decline/hold action. Each stage can be individually validated and behave exactly as specified, while the composed pipeline produces an outcome (a customer wrongly declined, a fraud ring undetected, a sanctioned entity processed) that no single-component audit would catch because the unit of analysis that fails is the pipeline, not any one vendor's model.

\paragraph{Why fraud pipelines are a harder composition case than MSRR.} Section~\ref{sec:l25msrr}'s case study treats vendor dependency as a performance-attribution problem: a vendor-side embedding update can shift the strategy's return distribution without tripping a single-component drift monitor. In a fraud pipeline, vendor composition carries an additional, non-substitutable exposure that has no analog in the trading case: regulated data scope. Several of the pipeline's intermediate stages process cardholder data, government ID numbers, or biometric templates directly, so composing a third-party LLM into any stage of this pipeline, for example, an LLM-based case-narrative summarizer or a conversational fraud-triage agent, raises a live PCI-DSS scope question: has cardholder data or its equivalent left the institution's compliance boundary the moment it is passed into a hosted third-party model's context window? This is a compliance exposure, not merely a performance-attribution one, and it does not appear anywhere in Section~\ref{sec:l25msrr}'s treatment of MSRR, since a news-embedding vendor never receives regulated customer data as input.

\paragraph{The composition failure this requirement is designed to catch.} Each stage of a fraud pipeline, audited in isolation, is unremarkable: the identity-verification vendor performs standard document and biometric matching as advertised; the device-fingerprinting vendor returns a standard risk signal; the LLM-based case-triage layer summarizes flagged transactions as specified. The composed pipeline, however, can produce a regulated-data exposure that no individual component audit would surface, because the failure is not in any one vendor's behavior but in an undocumented data flow across the interface between two correctly-functioning components: for instance, a case-triage LLM that was scoped to receive only a transaction ID and a risk score, but which a later integration silently extended to also pass the full transaction narrative, including an unmasked account number embedded in a free-text memo field, into a hosted third-party model's context window. This is structurally the same class of failure Section~\ref{sec:l25msrr} documents for MSRR's embedding-model dependency. An update or scope change to one component silently alters the pipeline's behavior without tripping any single-component monitor, but the consequence here is a regulatory data-scope breach rather than a shift in the realized Sharpe ratio, which is why the extended attestation in item~(5) above is proposed as a distinct, mandatory Layer~3 control for any fraud-decision-making pipeline with an agentic or foundation-model-dependent stage, rather than treated as an optional extension of the trading-specific requirement of Section~\ref{sec:l25msrr}.

\subsection{Layer 4: Systemic Layer (What You Contribute To)}\label{sec:l3}
Firm-level governance cannot mitigate risks that manifest only at the market level. When multiple firms train agents on similar, heavily overlapping data toward a similar Sharpe-style objective, position correlation is roughly stable across regimes, but joint drawdown risk rises sharply under stress, from 39.2\% to 79.3\% in our calm-vs-stress simulation. This change is caused by a common stress signal overwhelming small differences in independently learned policy parameters. Correlation across institutions need not be only mechanical. \citet{dou2025collusion} and \citet{dou2026planning} show that independently deployed reinforcement-learning and planning agents can learn collusive or bubble-riding strategies without agreement, communication, or intent, so a population of agents can become correlated through what each learns about the others, not only through what they all read. Layer~4's disclosure regime addresses both channels: the shared-input channel directly, through disclosure of common data and model dependencies, and the strategic channel indirectly, by giving supervisors the cross-institutional view needed to detect coordinated behavior that no single firm's monitor can see.

\subsubsection{Layer 4, Extended: Calibrating the Calm-versus-Stress Crowding Simulation}\label{sec:l3calib}
We construct a minimal structural model reproducing the calm-vs-stress joint drawdown figures reported in Section~\ref{sec:l3} (39.2\% calm, 79.3\% stress) to make explicit both the mechanism and the parameters underlying the claim.

\paragraph{Structural model.} Two agents' per-period returns are driven by a shared public factor plus idiosyncratic noise,
\[
r_{i,t}=\mu+\beta F_t+\sqrt{1-\beta^2}\,\varepsilon_{i,t},\qquad i\in\{1,2\},
\]
where $F_t\sim N(0,1)$ represents a shared public signal (e.g., a common news feed that both agents' strategies condition on), $\varepsilon_{i,t}\sim N(0,1)$ are independent idiosyncratic shocks, $\mu$ is a regime-level mean shift, and $\beta\in[0,1]$ is each agent's loading on the shared factor. Since $\mathrm{Corr}(r_1,r_2)=\beta^2$, $\beta$ is the model's crowding parameter: it captures the share of each agent's return explained by the same signal that other agents are also trading on. A drawdown is defined as $r_{i,t}<0$.

\paragraph{Calibration.} We solve for $(\mu,\beta)$ in each regime so that the analytic bivariate-normal joint-breach probability matches the target figures exactly, then confirm the match via Monte Carlo simulation ($n=10^6$ draws) of the structural model above; that is, the simulation is mechanistic (shared-factor loading), not merely a probability calculation. Individual (marginal) breach probability is set higher in the stress regime (0.85 vs.\ 0.60), reflecting that a genuine stress regime is, by construction, one in which the shared factor itself is realized adversely for a crowded position. Table~\ref{tab:crowd} reports the calibrated parameters and the resulting Monte Carlo match to target.

\begin{table}[ht]
\centering\small
\caption{Two-agent crowding simulation: calibration and results}\label{tab:crowd}
\begin{tabular}{lccccc}
\toprule
Regime & $\mu$ & $\beta$ & Implied corr.\ ($\beta^2$) & Target joint prob. & MC joint prob.\\
\midrule
Calm & $-0.253$ & 0.460 & 0.212 & 39.2\% & 39.1\%\\
Stress & $-1.036$ & 0.901 & 0.812 & 79.3\% & 79.2\%\\
\bottomrule
\end{tabular}
\end{table}

\begin{figure}[ht]
\centering
\includegraphics[width=0.95\textwidth]{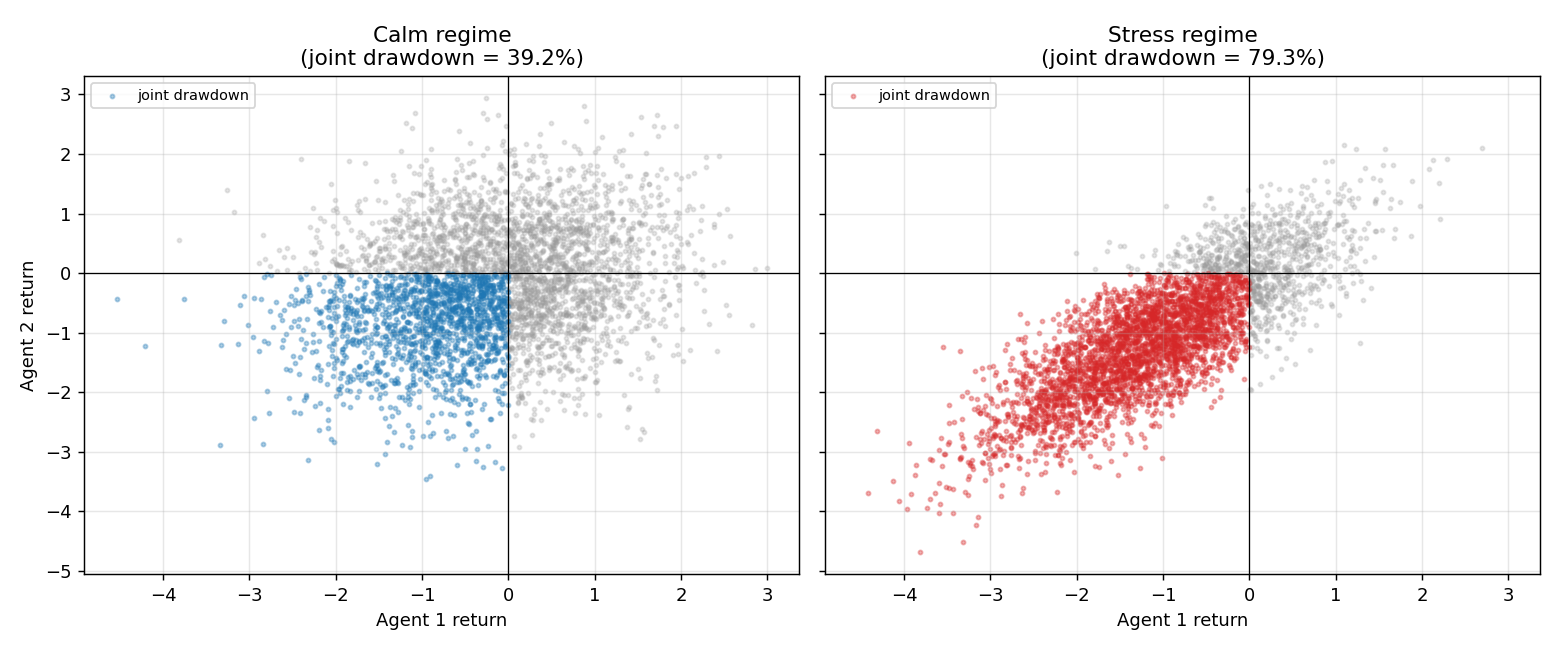}
\caption{Joint return distribution of two agents under the calm regime (left) and stress regime (right), $n=4{,}000$ simulated draws shown. Points in the lower-left (joint drawdown) quadrant are highlighted; the increase in density there reflects the rise in shared-factor loading $\beta$, not a change in either agent's marginal risk in isolation.}\label{fig:crowdscatter}
\end{figure}

\paragraph{Result.} Going from the calm to the stress regime, each agent's loading on the shared public signal rises from $\beta\approx0.46$ to $\beta\approx0.90$ (implied return correlation from $\approx0.21$ to $\approx0.81$), compounding with a regime-level mean shift to produce the reported jump in joint drawdown probability (Figure~\ref{fig:crowdscatter}). This is the mechanism described narratively in Section~\ref{sec:l3}: it is not that either agent's idiosyncratic behavior changes, but that a larger share of each agent's return becomes explained by the same shared signal. Training on a common public news feed toward similar objectives produces this result (Section~\ref{sec:msrrcrowd}).

\begin{figure}[ht]
\centering
\includegraphics[width=0.7\textwidth]{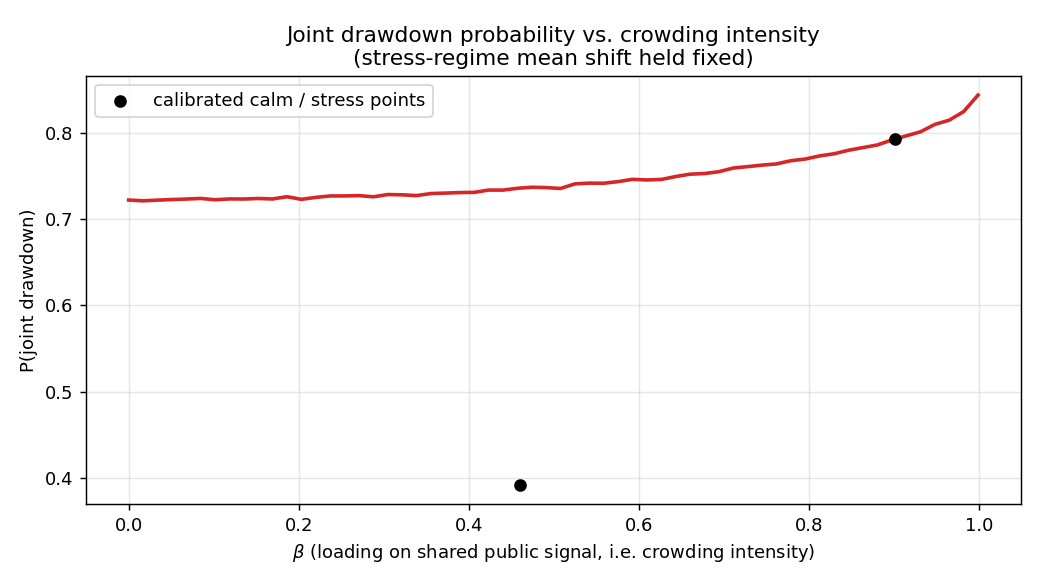}
\caption{Joint drawdown probability as a continuous function of the common-factor loading $\beta$ (stress-regime mean shift held fixed). The two calibrated calm/stress points from Table~\ref{tab:crowd} are marked.}\label{fig:crowdbeta}
\end{figure}

\paragraph{Sensitivity.} To show that the calibration is not a two-point artifact, Figure~\ref{fig:crowdbeta} sweeps $\beta$ continuously (holding the stress-regime mean shift fixed) and plots the resulting joint drawdown probability. Joint tail risk is sharply convex in crowding intensity: a moderate increase in shared signal reliance produces a disproportionate increase in systemic joint-loss probability. This probability is the empirical basis for Layer~4's disclosure requirement: a firm's own risk process cannot observe $\beta$ directly, since it depends on how many other institutions are exposed to the same shared signal.

As with Section~\ref{sec:l2drift}, this is a transparent, calibrated reconstruction of the reported figures under explicitly stated assumptions. This is consistent with our characterization of the Section~\ref{sec:l3} simulation as illustrative (Section~\ref{sec:limits}), rather than a claim to recover an unpublished original methodology.

\subsubsection{Layer 4, Extended: Crowding Risk from Shared Fraud-Consortium Data}\label{sec:l3fraud}
Section~\ref{sec:l3} formalizes crowding as a mechanism in which multiple firms training agents on similar, heavily overlapping data toward a similar objective produce a joint drawdown probability that rises sharply under stress, even though position correlation looks stable across regimes because a common stress signal overwhelms small differences in independently learned policy parameters. Section~\ref{sec:msrrcrowd} grounds this mechanism empirically in a shared public news feed. Fraud-decision-making agents are exposed to a structurally identical crowding channel, but through a data-sharing arrangement that is more direct, more concentrated, and in several respects more consequential than a shared public news feed: fraud consortium data.

\paragraph{The consortium-data channel.} Many financial institutions license or contribute to shared fraud-intelligence infrastructure: consortium negative lists, shared device-fingerprint graphs, shared identity-velocity signals aggregated across participating institutions, and, increasingly, a shared or small number of dominant third-party fraud-scoring vendors whose models are trained in part on this pooled data. This is a stronger form of the crowding channel Section~\ref{sec:msrrcrowd} documents for MSRR. A public news feed is symmetric and unconcentrated: every participant reads the same articles, but no single vendor controls the feed itself. Consortium fraud data is asymmetric and concentrated: a small number of consortium operators and fraud-scoring vendors sit structurally upstream of a large share of the industry's decision-making agents, so the shared-factor loading $\beta$ in Section~\ref{sec:l3calib}'s structural model is not merely an empirical regularity that emerges from independently-trained agents converging on similar objectives, as in the MSRR case, but is in part a designed property of the infrastructure itself: institutions adopt consortium data specifically because shared exposure improves individual detection performance, meaning the crowding channel is a deliberate, contracted dependency rather than an incidental one.

\paragraph{Restating the structural model for the fraud case.} Adapting Section~\ref{sec:l3calib}'s two-agent model directly, let the per-period fraud-decision-making outcomes of two institutions be driven by a shared consortium signal plus an idiosyncratic institution-specific signal,
\begin{equation}\label{eq:consortium}
d_{i,t}=\mu+\beta F_t+\sqrt{1-\beta^2}\,\varepsilon_{i,t},\qquad i\in\{1,2\},
\end{equation}
where $d_{i,t}$ is institution $i$'s decision-making outcome (e.g., a standardized measure of decline-rate error, false-positive, or false-negative, in period $t$), $F_t\sim N(0,1)$ represents the shared consortium signal (a pooled negative list, a shared device-graph risk score, or a common upstream vendor model), $\varepsilon_{i,t}$ is each institution's idiosyncratic detection signal, and $\beta$ is each institution's loading on the shared consortium data. Exactly as in Section~\ref{sec:l3calib}, $\mathrm{Corr}(d_1,d_2)=\beta^2$, and $\beta$ is not observable to any single institution's own risk process, since it depends on how many other consortium members are exposed to, and how heavily each weights, the same pooled signal.

A joint miscalibration event is defined here as $d_{i,t}$ crossing a harm threshold simultaneously across institutions, the fraud-domain analog of Section~\ref{sec:l3}'s joint drawdown, but with a materially different economic character: where Section~\ref{sec:l3}'s joint drawdown is a correlated loss event, the fraud-domain joint miscalibration event is more often a correlated false-positive spike, a shared consortium signal or vendor model update causing many institutions to simultaneously and wrongly decline or flag a common set of legitimate customers, a systemic customer-experience and reputational event rather than a P\&L event, and one with no clean analog in Section~\ref{sec:l3}'s trading-crowding formalization.

\paragraph{Why the fraud channel is structurally worse, not merely analogous.} Section~\ref{sec:l3calib}'s calibration shows the joint drawdown probability rising from 39.2\% to 79.3\% between calm and stress regimes as the shared-factor loading $\beta$ rises from $\approx0.46$ to $\approx0.90$. The consortium-data case plausibly exhibits a higher baseline $\beta$ even in calm regimes than the MSRR news-feed case, for three reasons specific to fraud infrastructure and absent from Section~\ref{sec:msrrcrowd}'s news-feed analysis:
\begin{enumerate}
\item \textbf{Contractual, not incidental, dependency.} Institutions opt into consortium data specifically to raise detection power, meaning the shared-factor loading is intentionally maximized by design, whereas news-feed crowding in Section~\ref{sec:msrrcrowd} is an emergent byproduct of independently-trained agents converging on similar public information.
\item \textbf{Vendor concentration.} Where the MSRR literature (Section~\ref{sec:msrrdesc}) documents strategy performance across a range of substitutable embedding vendors (BERT, GPT-2, Mistral-7B, Llama3 at multiple parameter counts), fraud-scoring infrastructure in payments is considerably more concentrated among a small number of dominant vendors. As a result, a single vendor-side model update, the same event Section~\ref{sec:l25fraud} identifies as a Layer~3 exposure, can propagate a correlated shift across a much larger share of the industry's decision-making agents than any single embedding-vendor update could propagate across trading strategies.
\item \textbf{Stress-regime correlation with a common trigger.} A genuine fraud-ring campaign or a large-scale data breach is, by construction, exactly the kind of shared adverse signal Section~\ref{sec:l3calib}'s stress regime models: it hits the shared consortium data feed for every participating institution simultaneously, which is also precisely the condition under which false-positive rates spike hardest, since models tuned to the shared signal react to the same anomalous input at the same time.
\end{enumerate}

\paragraph{Layer 4 disclosure requirement, extended.} Section~\ref{sec:msrrcrowd} proposes that the disclosure regime of Section~\ref{sec:l3} include, as a reportable field, the identity and provider of any third-party news or embedding feed on which a firm's agentic strategies depend. We extend this directly: the same disclosure regime should require an institution to report the identity of any fraud consortium, shared device-fingerprint graph, or third-party fraud-scoring vendor on which its customer-facing decision-making agents (Section~\ref{sec:l2bounded}) or transaction-level fraud-decision-making agents (Section~\ref{sec:l2kill}) depend, as well as the approximate share of the agent's decision-making weight attributable to that shared signal versus the institution-proprietary signal. This is a directly observable and, unlike Section~\ref{sec:l3calib}'s simulated $\beta$, directly measurable quantity in the fraud setting, since consortium data providers can, in principle, report participant counts, and the deploying institution can, in principle, measure feature importance attributable to consortium-sourced fields, giving Layer~4 a concrete instrumentation path for fraud pipelines that the news-feed case in Section~\ref{sec:msrrcrowd} could only argue for qualitatively.

\paragraph{Implication.} A firm evaluating its own fraud-decision-making agent in isolation, using only the extended Layer~2 controls of Sections~\ref{sec:l2kill} and~\ref{sec:l2bounded}, and the extended Layer~3 controls of Section~\ref{sec:l25fraud}, would correctly validate the model's decision boundary and bound its action authority, but would have no mechanism for detecting that a false-positive spike is about to hit its own customer base because it is about to hit every other consortium participant's customer base at the same moment, for the same upstream reason. This is the fraud-domain instance of the gap Section~\ref{sec:msrrcrowd}'s Layer~4 disclosure regime is designed to close, made sharper here by the fact that, unlike a public news feed, consortium participation is a named, contracted relationship that a firm's own governance process already has the visibility to disclose; it need only be required to.

\subsection{A Formal Agentic Risk Taxonomy}\label{sec:taxonomy}
We propose four ordinal dimensions, each scored 1--3: Autonomy ($A$), Reversibility ($R$), Blast Radius ($B$), and Policy Mutability Rate ($M$), combined into a composite Agentic Risk Score
\[
\mathrm{ARS}(s)=w_A A(s)+w_R R(s)+w_B B(s)+w_M M(s),\qquad \mathrm{ARS}(s)\in[1,3],
\]
with institution-specific weights $w_A,w_R,w_B,w_M$ summing to one and $\mathrm{ARS}(s)\geq2.5$ proposed as a provisional threshold for mandatory Layer~2 controls.

\subsection{Case Study: The LLM-Embedding News Anomaly as a Layer 3/4 Illustration}\label{sec:msrr}
Sections~\ref{sec:l25} and~\ref{sec:l3} motivate Layer~3 and Layer~4 with, respectively, a stylized multi-module pipeline and a minimal simulated two-agent market. \citet{didisheim2026} provide a real, economically large, and fully documented instance of exactly the system these layers are designed to govern, which we use here to make both layers concrete.

\begin{table}[ht]
\centering\small
\caption{Illustrative taxonomy scoring, extended with the news-anomaly case study (Section~\ref{sec:msrr})}\label{tab:tax_msrr}
\begin{tabular}{lcccc}
\toprule
System & $A$ & $R$ & $B$ & $M$\\
\midrule
Pre/intra/post-trade rules engine & 1--2 & 2 & 2 & 1\\
Single RL trading agent, online policy updates & 3 & 2--3 & 2 & 3\\
Multi-module content pipeline & 2--3 & 3 & 1--2 & 2\\
LLM-embedding news strategy (MSRR) & 1--2 & 2 & 2--3 & 2\\
\bottomrule
\end{tabular}
\end{table}

\subsubsection{System description and taxonomy scoring}\label{sec:msrrdesc}
\citet{didisheim2026} construct a long-short equity strategy (``MSRR'') that trades on the residual, or ``pure,'' component of LLM-derived news-article embeddings after purging the embeddings of content predictable from standard stock characteristics. The strategy is periodically retrained on an expanding window (an initial 24-month window, refit monthly thereafter) and achieves an out-of-sample Sharpe ratio of 3.1 over 1996--2022, which is roughly twice the strongest individual anomaly in the \citet{jensen2022} (JKP) universe. The strategy is also monotonically increasing in the parameter count of the underlying embedding model, rising from a Sharpe ratio of 1.5 with a 110-million-parameter BERT encoder to 4.1 with a 405-billion-parameter Llama3 model. Turnover is 75\%, exceeding every JKP factor.

\subsubsection{Layer 4: crowding as observed cross-institutional correlation}\label{sec:msrrcrowd}
Section~\ref{sec:l3calib}'s example is a toy simulation because no firm observes another firm's live agent. The news-anomaly literature offers an observable proxy for that unobserved correlation. \citet{didisheim2026} report that the rolling five-year Sharpe ratio of the pure news strategy, which ranged between roughly 2.1 and 4.5 over the full sample, drifts down from peaks near 4.5 to the 2.1--2.5 range in the years following the introduction of widely available transformer-based embedding models (post-2018), and attribute this decline to LLMs aiding the integration of news-based information into asset managers' portfolios more broadly, i.e., to crowding. This is the mechanism formalized in Section~\ref{sec:l3}: as more institutions deploy agents trained on the same public news feed toward closely related Sharpe-style objectives, the strategies' returns become more correlated in aggregate, even though no single firm observes the others' agents, and the market-level consequence (return decay, and by the logic of Section~\ref{sec:l3calib}, synchronized drawdown risk in stress regimes) is not visible to, or addressed by, any individual firm's governance process. A firm evaluating its own MSRR-style strategy in isolation, using only Layers~1 and~2, would correctly validate the reward function and the execution controls, but would have no mechanism for detecting that its strategy's true risk profile depends on how many other institutions are running a similar strategy against the same news feed. This is the gap Layer~4's policy-similarity disclosure regime is designed to close. We propose that the disclosure regime of Section~\ref{sec:l3} explicitly include as a reportable field the identity and provider of any third-party news or embedding feed on which a firm's agentic strategies depend. Shared upstream data (not merely a shared reward objective) is an independent and observable channel of cross-institutional correlation that this case study makes newly visible.

\subsubsection{Implications}\label{sec:msrrimpl}
Taken together, the news-anomaly case study demonstrates that Layers~3 and~4 are not just speculative extensions motivated only by hypothetical multi-agent pipelines. The strategy class that is already deployed, economically large (Sharpe ratios exceeding 3, net of realistic trading costs after lookback smoothing), dependent on opaque third-party foundation models, and exhibiting the return decay signature of crowding has positioned itself inside the taxonomy's $M=2$, $B=2$--3 region this paper identifies as requiring mandatory Layer~2 and Layer~3 controls under a $w$-weighting that gives any material weight to vendor dependency or blast radius.

\subsection{Case Study: The Situational Awareness Blowup as a Layer 1/4 Illustration, and the Limits of Layer 2}\label{sec:sa}
Section~\ref{sec:msrr} positioned Layers~3 and~4 in an agentic pipeline. Here, we consider a contemporaneous event that impacts Layers~1 and~4 from a different angle. It is instructive because the process shows where this paper's Layer~2 controls do not apply, which sharpens the case for treating Layer~2 as a distinct, non-substitutable layer.

\subsubsection{System description}\label{sec:sadesc}
Situational Awareness LP, founded in 2024 by former OpenAI researcher Leopold Aschenbrenner, is a discretionary hedge fund built around a single thesis: that the continued scaling of compute and algorithmic efficiency makes AGI by approximately 2027 sufficiently plausible to justify concentrated, leveraged exposure to the AI infrastructure buildout---long chips, memory, data centers, and power; short software incumbents viewed as vulnerable to AI-driven disruption.\footnote{Disruption Banking, July 30, 2026.} The fund is not an agentic or algorithmic trading system in the sense governed by Layers~1--3 of this paper: positions are set by human portfolio managers on a discretionary, thesis-driven basis, not generated by a retrained policy or an autonomous agent. This distinction matters for what follows.

By early July 2026, the fund had grown to a reported peak of roughly \$45 billion in assets under management, up approximately 450\% year-to-date, with a \emph{Financial Times}-reviewed investor letter reporting a 439\% net return for H1 2026.\footnote{Disruption Banking, July 30, 2026, citing \emph{Financial Times} investor letter dated July 24, 2026.} Reported leverage on the public book ran as high as 4x (400\%).\footnote{CNBC, July 31, 2026; social-media reporting on the fund's timeline.} A sharp reversal in AI infrastructure names began around July 10 and was variously described as triggered or signaled by the SK Hynix U.S.\ IPO. The reversal produced a two-week selloff in which many AI-adjacent equities fell 30\% or more.\footnote{CNBC, July 31, 2026.} The leveraged book generated margin calls on both the long and short legs simultaneously, and the fund was forced to sell its entire public equity portfolio in a block trade to Citadel at a discount; reported AUM fell from \$45 billion to approximately \$10 billion within weeks.\footnote{Bloomberg, July 30, 2026; CNBC, July 30--31, 2026.} The fund retained its unlevered and private holdings (including stakes in Anthropic and other AI infrastructure names) and remained up substantially year-to-date despite the forced sale.\footnote{CNBC, July 31, 2026.}

\subsubsection{Taxonomy scoring, and why it differs in kind from the MSRR case}\label{sec:satax}
Scored against the taxonomy of Section~\ref{sec:taxonomy}, Situational Awareness sits almost nowhere near the MSRR strategy of Section~\ref{sec:msrr}, and the difference is the point:

\begin{table}[ht]
\centering\small
\caption{Taxonomy scoring extended with the Situational Awareness case study}\label{tab:tax_sa}
\begin{tabular}{lcccc}
\toprule
System & $A$ & $R$ & $B$ & $M$\\
\midrule
Pre/intra/post-trade rules engine & 1--2 & 2 & 2 & 1\\
Single RL trading agent, online policy updates & 3 & 2--3 & 2 & 3\\
LLM-embedding news strategy (MSRR) & 1--2 & 2 & 2--3 & 2\\
Situational Awareness (discretionary, levered thesis fund) & 1 & 2--3 & 3 & 1\\
Real-time payments fraud-decision agent & 3 & 1 & 2--3 & 2--3\\
Customer-facing conversational/action agent & 2--3 & 1--2 & 2 & 2\\
\bottomrule
\end{tabular}
\end{table}

The payments fraud-decision agent scores high on autonomy ($A=3$): decline and hold decisions are executed directly against a live transaction, with no human in the loop at the point of action. Reversibility is low ($R=1$): once a legitimate transaction is declined, the direct harm to the customer relationship, and in some cases to the transaction itself, cannot be undone by later correcting the model, distinguishing it sharply from the trading agents in Table~\ref{tab:tax_sa}, where a mispriced position can still be unwound at a cost. The blast radius is firm-wide in the base case and market-adjacent under stress ($B=2$--3): a single miscalibrated model degrades one institution's decline rate, but a shared vendor fraud model or shared consortium data feed (Section~\ref{sec:l3} crowding mechanism, restated for payments) can produce a correlated spike in false declines across multiple institutions simultaneously. Mutability is periodic to near-continuous ($M=2$--3), since many production fraud models retrain on a rolling window to track adversarial drift from fraud rings actively probing the decision boundary, a faster and more adversarial retraining cadence than MSRR's monthly refit (Section~\ref{sec:msrrdesc}).

The customer-facing conversational/action agent scores moderate-to-high on autonomy ($A=2$--3): $A=3$ once the agent is authorized to take binding actions (refunds, credit changes, dispute resolution) rather than only generating text for a human to approve, the same autonomy escalation that applies to MSRR once its output is wired directly into execution. Reversibility is impaired ($R=1$--2): a commitment made to a customer, whether contractual or reputational, is costly to retract even when the underlying account action can technically be reversed. Blast radius is firm-wide ($B=2$): harm is typically per-customer rather than systemic, absent a shared foundation-model vendor failure affecting many institutions' deployments at once. Mutability is periodic ($M=2$), tracking the underlying LLM's update cadence, which the deploying institution frequently does not control, directly paralleling the vendor-dependency argument of Section~\ref{sec:l25msrr}.

In contrast, the hedge fund Situational Awareness is scored low on autonomy ($A=1$): a human sets and can override every position. Mutability is also low ($M=1$): the thesis itself did not update in production the way a retrained policy does. For this hedge fund, the problem was a thesis that was held too steadily through a regime change, not one that drifted unpredictably. Blast radius is high ($B=3$): a \$45 billion book with 4x leverage and a single-factor concentration (the AI-capex-continues trade) creates firm- and market-relevant loss potential, regardless of how the positions were generated. Reversibility is impaired ($R=2$--3) not because the strategy was opaque, but because forced deleveraging under margin calls compresses the exit price below any orderly-unwind estimate. The realized loss is a function of when you are forced to sell, not just what you hold.

The practical upshot: an ARS-style score built only from $(A,M)$, autonomy and mutability, the two dimensions this paper's Layer~2 is most directly built to address, would significantly under-flag this fund, making it seem safer than the MSRR pipeline. It was not. This is direct evidence that ARS's four dimensions are not interchangeable and that a firm (or an investor doing due diligence) that governs only for autonomy and policy drift, while leaving blast radius and leverage-driven reversibility ungoverned, has a real gap.

\subsubsection{Layer 1 Discussion: the thesis as an ungoverned reward function}\label{sec:sal1}
Section~\ref{sec:l1} argues that a reward function is the mathematical expression of institutional intent and should receive policy-document-grade review. The Situational Awareness thesis functioned as exactly this kind of governing objective for a \$45 billion book, without the formal review Layer~1 proposes for an agentic system's reward function. Three questions Layer~1 would have forced onto the table, and that a reward-function-style review checklist would surface for a discretionary thesis fund as readily as for an RL agent:
\begin{itemize}
\item \textbf{What is the objective actually optimizing, and under what regime does it fail?} ``AGI by $\sim$2027 justifies concentrated AI-infrastructure exposure'' is a directional thesis, not a risk-bounded objective. It says nothing about position sizing, leverage limits, or what evidence would falsify it in the short-to-medium term versus the long term. This is the same gap that Layer~1 identifies in an unreviewed reward function that specifies a goal but not the guardrails around pursuing it.
\item \textbf{Is the objective's time horizon mismatched to the vehicle's liquidity structure?} The thesis is a multi-year structural call; the vehicle carried 4x leverage in liquid public equities, which is a short-horizon tolerance instrument. A reward-function review process, applied by analogy, would ask whether the implementation (leverage, concentration, liquid public names) is consistent with the objective's actual time horizon, flagging the mismatch that produced the forced sale.
\item \textbf{Who signed off on leverage as an amplifier of the thesis, separate from signing off on the thesis itself?} This is the discretionary-fund analog of Layer~1's requirement that the reward function receive independent review distinct from the strategy that pursues it.
\end{itemize}

\subsubsection{Layer 2 Discussion: why the engineering layer's specific controls do not transfer here and why that is informative}\label{sec:sal2}
It would be a stretch to claim that policy-stability monitoring or an inner-confidence kill-switch (Section~\ref{sec:l2}) ``would have prevented'' this blowup because there was no model policy to monitor and no LLM decision confidence generating the trades. Claiming otherwise would be exactly the kind of category error this paper spends Section~\ref{sec:fail} warning against: applying an agentic-system control to a system that is not agentic.

What transfers is the underlying design principle behind Layer~2, restated at the right level of abstraction: \textbf{a governed system needs a mechanism that can force de-risking before the system's own dynamics do so involuntarily, at a worse price.} For an RL policy, that mechanism is a kill-switch on inner confidence. For a leveraged discretionary book, the equivalent mechanism is a pre-committed, leverage-tiered de-risking schedule (e.g., automatic deleveraging at defined drawdown or margin-utilization thresholds, set and reviewed independently of the portfolio manager, before a crisis). Reporting indicates that Situational Awareness's forced sale was a margin-call-driven event, not a voluntary risk-reduction one. That is, de-risking arrived through the market's mechanism (the prime broker's margin desk) rather than through the fund's own. This is the discretionary-fund equivalent of a kill-switch never firing until an external actor pulls it. The lesson for governance design is that Layer~2 is not ``the LLM-specific layer'' so much as ``the layer where a forcing function for de-risking is pre-committed rather than improvised'', a principle that generalizes even to systems with no model in the loop, though its implementation does not.

\subsubsection{Layer 4 Discussion: crowding, observed directly}\label{sec:sal3}
This case is a cleaner, faster-moving instance of the crowding mechanism Section~\ref{sec:l3} models and Section~\ref{sec:msrrcrowd} evidences via a five-year Sharpe drift. Situational Awareness's long book was concentrated in widely-held names such as chip, memory, data-center, and power supplier companies. These names were, by 2026, a widely-held consensus trade across the market, not a proprietary position. The reversal was sharp because many participants were positioned the same way: when infrastructure-adjacent names turned, the unwind was not idiosyncratic to one fund's book. It was correlated selling across a crowded trade, which is what produced 30\%+ two-week moves in ``most AI stocks,'' not just Situational Awareness's holdings. A market commentator characterizing the forced sale as potentially ``a clearing event'' for the broader AI trade is, in this paper's terms, an informal statement of exactly the Layer~4 mechanism: firm-level risk that is only legible at the market level because it depends on how many other participants hold the same exposure.

Layer~4's proposed disclosure regime (Section~\ref{sec:l3}) would not have prevented the reversal: crowding is a market-structure fact, not a single firm's governance failure. However, Layer~4's regime would have given both the fund's own risk function and its counterparties (prime brokers extending 4x leverage) better information about how correlated the book's true risk was with the rest of the market's AI-infrastructure exposure, which bears directly on whether 4x leverage was ever an appropriate facility to extend against a single-factor, crowded trade.

\subsubsection{Implications}\label{sec:saimpl}
The two case studies in this paper are complementary rather than redundant. Section~\ref{sec:msrr}'s MSRR pipeline shows what governance failure looks like when the system is agentic but low-autonomy on paper ($A=1$--2): the risk hides in composition and vendor dependency, which is why Layer~3 exists. Section~\ref{sec:sa}'s Situational Awareness case shows what governance failure looks like when the system has no agentic component at all ($A=1$) but scores maximally on blast radius through leverage and crowding. This is why Layer~4 and a leverage-aware reading of Layer~1 matter even for a firm with nothing ``agentic'' in production. A firm that reads this paper as ``a framework for AI trading agents'' and concludes it has no application because its book is human-managed would be making the precise category error Section~\ref{sec:sa} warns against: the specific engineering controls of Layer~2 are model-specific, but the underlying governance discipline (reward/thesis review, pre-committed de-risking, and crowding disclosure) is not. This case is evidence that the gap it closes is not hypothetical.

\subsection{Discretionary Risk vs.\ Governed Algorithmic Risk: Why ``AI-Native'' Is Not Automatically Safer, But Can Be Made Structurally Safer}\label{sec:disc}
Section~\ref{sec:sa} documented a discretionary fund whose risk was entirely based on human judgment, leverage, and thesis. Nothing about the fund's failure was specific to AI. A natural question follows: would a genuinely algorithmic, model-driven fund, one whose research inputs and trading decisions are generated by a governed system, have been structurally less prone to this kind of blowup? We argue yes, but only conditionally, and the condition is precisely the governance apparatus this paper proposes. An ungoverned algorithmic fund is not safer than Situational Awareness; it is differently dangerous, in ways this paper's own taxonomy predicts.

\subsubsection{The structural case for AI-native governance}\label{sec:disc1}
A true AI-native trading system is one where position sizing, leverage, and risk limits are enforced in code rather than held as intentions. When built as such, a true AI-native trading system has three advantages that a discretionary fund structurally lacks, corresponding directly to Layers~1, 2, and~3 of this framework:

\textbf{Risk limits are enforced, not just observed.} In a discretionary fund, a leverage cap is a policy that a human is supposed to respect under stress. Unfortunately, this is the condition under which humans are least reliable, and exactly the condition under which Situational Awareness's book was not deleveraged until a prime broker's margin desk forced the issue.

In contrast, in a governed algorithmic system, Layer~2's kill-switch architecture (Section~\ref{sec:l2}) can be wired directly into the execution path: a leverage or drawdown threshold breach halts new position-taking or triggers automatic deleveraging as a hard constraint, not a discretionary call made in the middle of a panic by the person with the most career and psychological investment in the thesis being right. This is the sharpest version of the argument: code does not get emotionally attached to a thesis it wrote, but a founder-PM can.

\textbf{The reward function can be inspected and reviewed before capital is at risk.} Layer~1 requires that the objective a system optimizes for be written down and reviewed with the rigor of a risk policy before deployment. A discretionary thesis (``AGI by $\sim$2027 justifies concentrated leveraged exposure'') is rarely forced through this discipline. The thesis resides in an investor letter and the PM's conviction, not in a document with a defined failure condition, position-sizing rule, and independent sign-off. An algorithmic system's objective function has to be written down by construction; it can be reviewed, red-teamed, and stress-tested against the adversarial input distributions Section~\ref{sec:adversarial} describes in a way that a human's evolving conviction cannot.

\textbf{Drift is measurable.} Section~\ref{sec:l2}'s integrated-covariance identity gives a governed algorithmic system an exact, continuously computable measure of how much of its objective depends on cost information, together with a passive upper bound that needs nothing but observed costs and decisions. Whether live behavior has diverged from validated behavior becomes a comparison of that measure with its validated level. A discretionary fund has no equivalent instrument for detecting that a PM's decision-making has quietly shifted. The fund can become more concentrated, more levered, more convinced by disconfirming evidence rather than updated by it, until the P\&L reveals it, which is the most expensive possible place to detect drift.

On this reading, an AI-native fund with Layers~1 and~2 actually implemented has a real structural edge: its risk controls are properties of the system's architecture rather than promises about the operator's future behavior under stress.

\subsubsection{The counter-case: this paper's own evidence that ``AI-native'' is not sufficient}\label{sec:disc2}
The argument above describes a governed algorithmic fund. This paper's central empirical claim (Section~\ref{sec:survey}) is that governance of exactly this kind is rare: 88\% of surveyed finance professionals report no operational governance framework for agentic AI, and only 32\% of money managers disclosing AI use in Form ADV filings disclose a formal governance policy. An algorithmic fund without Layers~1--3 in place does not inherit the structural advantages of Section~\ref{sec:disc1} by virtue of being algorithmic. A fund without these risk control layers inherits a different, and in some respects less visible, set of failure modes:
\begin{itemize}
\item \textbf{Policy drift is silent where discretionary drift is at least eventually legible.} A PM's growing overconfidence shows up in position sizing that a risk committee can observe and question. A retraining agent's policy drift (Section~\ref{sec:aisafety}) can shift the system's effective risk-taking without any human decision that anyone could have vetoed. There was no moment analogous to Aschenbrenner choosing 4x leverage that a compliance officer could have flagged because no human chose the updated policy.
\item \textbf{Compositional risk (Layer 3) has no discretionary analog at all.} Situational Awareness's failure was legible in principle to any experienced risk officer: concentrated, leveraged, single-factor exposure is a recognizable pattern. The MSRR case study (Section~\ref{sec:msrr}) shows a failure mode in which a vendor-side embedding model update silently reshapes a pipeline's return distribution. This failure mode is not recognizable to conventional risk review because every individual component looks fine in isolation. An AI-native fund is more exposed to this class of failure than a discretionary one, not less, precisely because it depends on components that a human trader's judgment never had to depend on.
\item \textbf{Crowding may be worse, not better, among algorithmic strategies.} Section~\ref{sec:msrrcrowd} provides evidence that Sharpe ratios on the LLM-embedding news strategy were compressing from the 4+ range toward 2--2.5 as the underlying embedding technology became widely available. This shows that algorithmic strategies trained on similar data toward similar objectives converge in a way discretionary managers, whose theses and information sources are more heterogeneous, may not. If every AI-native fund's risk model is built on similar foundation-model embeddings and similar training objectives, Layer~4 crowding risk could be structurally worse for a population of AI-native funds than for a population of discretionary ones. Correlated model architecture is a new, additional channel of correlation beyond correlated conviction.
\item \textbf{A governed system can enforce the wrong policy just as reliably as the right one.} Layer~2's advantage is that risk limits are enforced in code, not merely observed, and it cuts both ways. An algorithmic system with a flawed or unreviewed reward function will pursue that flawed objective with the same mechanical reliability it would apply to a sound one. Layer~1 review is what makes hard-coded enforcement an advantage rather than a liability; absent it, ``the code does not get emotionally attached to its thesis'' becomes ``the code will drive the position to zero without a moment of doubt,'' which is a worse property, not a better one, in the specific case where the thesis is wrong. The same holds for the drift monitor: it measures departure from the validated policy, and so it certifies a flawed policy as stable as readily as a sound one.
\end{itemize}

\subsubsection{The synthesis: the advantage is real but earned, not structural}\label{sec:disc3}
The defensible version of the claim is therefore: \textbf{a fund whose trading decisions are generated by a system with Layers~1--3 actually implemented has risk controls a discretionary fund cannot match, because those controls are architectural rather than behavioral.} The unqualified version of ``AI hedge funds are inherently sounder than discretionary ones'' is not supported by this paper's own evidence, since it is precisely the population of agentic financial systems this paper studies that shows an 88\% governance gap and a real, currently-observed compositional and crowding failure mode (Section~\ref{sec:msrr}) that has no discretionary equivalent.

Put differently, the failure of Situational Awareness was a known, well-understood risk management problem (leverage against a concentrated, crowded trade) that traditional risk management has fifty years of tools to address, deployed too weakly or too late. A poorly-governed AI-native fund's failure mode is a silent policy drift compounded by opaque vendor dependency, which is further compounded by algorithmic crowding invisible to any single firm's risk process. This is a novel risk-management problem that this paper argues the industry does not yet have the tools, or in 88\% of cases the process, to address at all. Whether AI-native funds are ``more sound'' than discretionary ones is therefore an empirical question that resolves in favor of AI-native funds only after the point where Layers~1--3 are actually built, and resolves against them, in a harder-to-detect way, on every day before that point.

\subsubsection{Implication for the paper's argument}\label{sec:disc4}
This case strengthens rather than weakens the paper's core thesis in Section~\ref{sec:intro}: the risk is not AI-native trading itself, and the answer is not ``prefer discretionary managers.'' The answer is that the structural advantage of AI-native systems is their capability for enforced limits, inspectable objectives, and measurable drift. The advantage is not automatic; it is unlocked only by the governance architecture this paper proposes. A firm that adopts an AI-native strategy without adopting Layers~1--3 has not reduced its exposure to a Situational-Awareness-style blowup; it has traded a visible, well-understood failure mode for a set of less visible ones that this paper's evidence suggests the industry is not yet equipped to see coming.

\subsection{Case Study Four: The Air Canada Chatbot Ruling as a Layer 1 Illustration, and the Limits of Component-Level Correctness}\label{sec:aircanada}
Sections~\ref{sec:msrr} and~\ref{sec:sa} use a deployed trading pipeline and a discretionary fund blowup to ground Layers~3 and~4, and Layers~1 and~4, respectively, in real, currently-documented events rather than only hypothetical systems. We add a third case study, grounded in a real, adjudicated customer-facing agentic incident, to make the bounded-action-authority control of Section~\ref{sec:l2bounded} and the cost-matrix framing of Layer~1 in Section~\ref{sec:l1fraud} concrete in exactly the way Section~\ref{sec:msrr} makes Layers~3 and~4 concrete for a trading pipeline.

\paragraph{System description.} In November 2022, a passenger named Jake Moffatt used Air Canada's website chatbot to ask about bereavement fares following the death of a family member. The chatbot responded that a bereavement discount could be claimed retroactively, within 90 days of ticket issuance, by submitting a refund application, and provided a link to the airline's official bereavement-fares policy page as part of its response.\footnote{\emph{Moffatt v.\ Air Canada}, 2024 BCCRT 149 (Civ.\ Resol.\ Trib.).} The linked policy page, however, stated the airline's actual policy: bereavement fares must be requested before travel and are not available retroactively. Relying on the chatbot's statement rather than the contradicting policy page to which it linked, Moffatt purchased full-fare tickets and later applied for the retroactive discount, which Air Canada denied, citing the correct policy. Moffatt brought the matter before the British Columbia Civil Resolution Tribunal. Air Canada's defense argued that the chatbot was, in effect, ``a separate legal entity that is responsible for its own actions,'' and that the airline should not be held liable for the chatbot's statements.\footnote{CBC News, February 16, 2024; \emph{Forbes}, February 19, 2024.} Tribunal Member Christopher Rivers rejected this argument, holding that Air Canada ``did not take reasonable care to ensure its chatbot was accurate'' and that the airline, not a separate entity, was responsible for all information on its website regardless of whether a human or an automated agent produced it. The tribunal awarded Moffatt \$812.02 in damages, interest, and fees, a small dollar figure, but a governance-relevant liability finding: the tribunal explicitly declined to treat the chatbot's output as categorically different from a statement made by a human representative of the airline.

\paragraph{Taxonomy scoring.} Scored against the taxonomy of Section~\ref{sec:taxonomy}, extended with the enterprise rows of Table~\ref{tab:tax_sa}:

\begin{table}[ht]
\centering\small
\caption{Taxonomy scoring extended with the Air Canada chatbot case study}\label{tab:tax_ac}
\begin{tabular}{lcccc}
\toprule
System & $A$ & $R$ & $B$ & $M$\\
\midrule
LLM-embedding news strategy (MSRR) & 1--2 & 2 & 2--3 & 2\\
Situational Awareness (discretionary, levered thesis fund) & 1 & 2--3 & 3 & 1\\
Customer-facing conversational/action agent (general) & 2--3 & 1--2 & 2 & 2\\
Air Canada bereavement-fare chatbot & 2 & 1 & 1--2 & 1--2\\
\bottomrule
\end{tabular}
\end{table}

Autonomy is moderate rather than high ($A=2$): the chatbot did not execute a binding transaction directly. However, the chatbot did make a representation that induced the customer to take an action (purchasing a full-fare ticket) in reliance on that representation. This is a materially different autonomy profile from the general customer-facing agent row, which contemplates an agent with direct action authority (Section~\ref{sec:l2bounded}). This is also why this case is instructive: the tribunal imposed liability on Air Canada despite the chatbot never having crossed any bounded action-authority threshold in the sense of Section~\ref{sec:l2bounded}, since it took no action at all; the chatbot only spoke. Reversibility is low ($R=1$): once the customer relied on the representation and purchased a ticket at full fare, the harm was not contingent on any subsequent system action and could not be undone by the airline correcting the chatbot after the fact. However, Air Canada only noted and corrected the issue internally, without remedying the affected customer's already-incurred loss. Blast radius is low-to-moderate in this single instance ($B=1$--2). However, this is a firm-wide exposure in aggregate, since the same failure mode (an unreviewed policy-page inconsistency) applies to every customer who queried the chatbot on the same topic, not only to Moffatt. Mutability is low ($M=1$--2): the chatbot's incorrect statement was not the product of a retrained policy drifting from a validated baseline in the sense of Section~\ref{sec:l2fraud}'s integrated-covariance mechanism; it was, per subsequent reporting, an unresolved inconsistency between the chatbot's underlying retrieval content and the airline's policy page that had likely existed since deployment, a static misalignment rather than a drifted one. A drift monitor calibrated on this chatbot's own behavior would therefore have certified it as stable, which is why Layer~1 review, not Layer~2 monitoring, is the control this case calls for.

\paragraph{Layer 1: an unreviewed representation-generating objective.} Section~\ref{sec:sal1} argues that the Situational Awareness thesis functioned as an ungoverned reward function for a \$45 billion book. The Air Canada chatbot's underlying retrieval and generation configuration functioned analogously, as an ungoverned objective for a system authorized to make representations that bind the institution, without the review Layer~1 proposes. The tribunal's reasoning makes this explicit: the airline's defense rested on treating the chatbot's output as categorically unreviewable company speech, precisely the compliance-as-PDF failure Section~\ref{sec:pdf} describes. An accurate written bereavement-fares policy existed, but nothing in the chatbot's execution path enforced consistency with that policy document. Framed in the cost-matrix terms of Section~\ref{sec:l1fraud}, the chatbot's implicit objective (to generate a plausible, helpful-sounding answer) was never reviewed against the asymmetric cost of a confidently wrong answer relative to a correctly declined-to-answer response, the same false-positive/false-negative asymmetry Section~\ref{sec:l1fraud} formalizes for fraud decision-making, here mapped onto ``answer confidently'' versus ``defer to human or refuse to answer'' rather than ``approve'' versus ``decline.''

\paragraph{Relation to bounded action authority.} A natural objection to treating this case as evidence for Section~\ref{sec:l2bounded}'s bounded-action-authority control is that the chatbot took no bounded action at all; it made a statement, not a transaction, so an action ceiling $\kappa(s)$ would not have applied. This objection is itself the case study's governance lesson: Section~\ref{sec:l2bounded}'s control, as specified, bounds the consequences of an agent's executed actions, and a purely representational agent that induces a customer to take their own action downstream of an unreviewed statement falls outside that boundary as written. We therefore extend Section~\ref{sec:l2bounded}'s framework: the consequence magnitude $v(a)$ used to calibrate an action ceiling should be defined to include reasonably foreseeable customer-initiated actions taken in reliance on an agent's representation, not only actions the agent itself executes, since the tribunal's negligent-misrepresentation finding did not turn on who clicked the purchase button; it turned on whether the airline took reasonable care to ensure that the chatbot's statements were accurate. A governance framework that bounds only executed actions and treats representations as outside its scope would, on this reading, have missed the exact failure this case documents.

\paragraph{Implications.} This case complements Sections~\ref{sec:msrr} and~\ref{sec:sa} by completing a third quadrant the paper's taxonomy predicts but has not yet been evidenced with a real incident: Section~\ref{sec:msrr}'s MSRR pipeline shows failure where the system is agentic but low-autonomy on paper. Section~\ref{sec:sa}'s Situational Awareness shows a failure where the system has no agentic component but scores maximally on blast radius. This case shows a failure where the system is customer-facing and low-consequence-per-instance in isolation; yet, it was found to carry full institutional liability precisely because no Layer~1-style review was ever applied to the objective generating its representations. The ruling's core holding is that an institution cannot disclaim responsibility for an agent's output by characterizing the agent as autonomous or separate from the institution. It is the clearest available real-world precedent for this paper's Section~\ref{sec:l1} argument: a reward function, or in this domain, a customer-facing agent's governing objective, is a governance artifact that the deploying institution owns, not a property of the agent that shields the institution once deployed.

\section{Implementation Roadmap: From Awareness to Governance in 90 Days}\label{sec:roadmap}
\subsection{Trading Track}\label{sec:roadtrading}
\textbf{Days 1--30:} Inventory all agentic systems, including embedding-model-dependent strategies; document vendor model identifiers per Section~\ref{sec:l25}, item~5; and record, for each system, which tier of the Layer~2 monitor (Table~\ref{tab:tiers}) its current access supports. \textbf{Days 31--60:} Implement policy stability monitoring and kill-switch architecture. Policy stability monitoring starts with the passive covariance bound for every inventoried system, which needs only observed costs and decisions, and moves to the exact integrated-covariance monitor (Algorithm~\ref{alg:monitor}) wherever the decision layer can be replayed: always for the institution's own portfolio-construction step, and in closed form when that step is affine (Section~\ref{sec:l25msrr}). Extend Layer~3 red-teaming, vendor-version attestation and replay-access contract terms (Section~\ref{sec:l25}, item~6) to any pipeline identified in the inventory that depends on a third-party LLM, foundation model, or embedding service. \textbf{Days 61--90:} Submit the reward function governance to model risk management; assess policy-similarity and shared data exposure to likely competitor deployments; present the systemic-layer assessment, including any material dependence on widely-used public news or embedding providers, to the CRO and board risk committee.

\subsection{Payments Track}\label{sec:roadpay}
\paragraph{Days 1--30: Inventory and Regulatory Mapping.} Inventory every agentic or foundation-model-dependent touchpoint with either (a) binding action authority over a customer account or transaction, or (b) the ability to generate a representation a customer could reasonably rely on, extending Section~\ref{sec:roadtrading}'s inventory step beyond trading systems to fraud-decision-making agents, customer-service chatbots, and any dispute-resolution or refund automation. For each system identified, document: the vendor and model identifier for any third-party foundation-model dependency, per the attestation requirement of Section~\ref{sec:l25fraud}; whether the system's outputs or actions fall under Regulation~E's error-resolution clock, BSA/AML's SAR-filing clock, or PSD2/PSD3's liability-allocation regime, per Section~\ref{sec:payreg}; and whether the system currently has any documented ceiling on the consequence magnitude of a unilateral action or representation, per Section~\ref{sec:l2bounded}. A system with no such ceiling documented should be treated, for the remainder of this 90-day track, as carrying an unbounded $\kappa(s)$ by default, not an unknown one.

\paragraph{Days 31--60: Engineering Controls and Cost-Matrix Review.} Implement the inner-confidence kill-switch of Section~\ref{sec:l2kill} and the integrated-covariance drift monitor of Section~\ref{sec:l2fraud}, with the regret-to-covariance ratio $\hat\rho_t$ as its primary statistic and segment attribution enabled, for every fraud-decision-making agent identified in the Days~1--30 inventory, with the calibration and recalibration cadence set to reflect the shorter, adversarially-adaptive horizon argued for in Section~\ref{sec:l2kill} rather than the longer horizon appropriate to a trading policy, and with the drift monitor's chargeback lag covered by the kill-switch. In parallel, implement the bounded action-authority ceilings $\kappa(s)$ and $K(s)$ of Section~\ref{sec:l2bounded} for every customer-facing agent identified, routing any action or representation above the ceiling to human review by default rather than to automatic denial or automatic execution. Submit the cost matrix of every fraud-decision-making agent, $c_{FP}$ and $c_{FN}$ as defined in Section~\ref{sec:l1fraud}, to the same independent review this step requires of any trading reward function, together with the disparate-impact analysis Section~\ref{sec:l1fraud} specifies as a precondition of that review, not a follow-on step. Because the drift monitor's regret is denominated in this cost matrix, the review should be completed before the monitor's trigger levels are calibrated.

\paragraph{Days 61--90: Composition Attestation, Consortium Disclosure, and Escalation.} Extend Layer~3 interface contracts and exclusion lists (Section~\ref{sec:l25fraud}) to every pipeline stage identified in the Days~1--30 inventory that touches cardholder data, government identifiers, or biometric templates, with particular attention to any stage where a foundation-model call was added to an existing pipeline after its original compliance scoping, the composition failure Section~\ref{sec:l25fraud} identifies as most likely to be missed by a routine audit. Submit, alongside the reward-function and policy-similarity assessments Section~\ref{sec:roadtrading} already requires for the CRO and board risk committee, the identity of any fraud consortium, shared device-fingerprint provider, or concentrated upstream fraud-scoring vendor that each agentic system depends on, together with the estimated share of decision-making weight attributable to that shared signal, per the disclosure extension of Section~\ref{sec:l3fraud}; and a mapping from each agentic touchpoint identified in Days~1--30 to the specific regulatory clock it is subject to (Reg~E's investigation and resolution windows, BSA/AML's 30/60-day SAR filing window, or PSD3's fraud-liability allocation triggers), so that Layer~2 escalation paths are designed against these external deadlines from the outset rather than discovered to conflict with them during an actual dispute or filing.

\paragraph{Relationship to the trading track.} The two tracks share a common Day~1--90 calendar and a common escalation endpoint, the CRO and board risk committee, but are intentionally decoupled in their Days~31--60 engineering work, since a trading desk's policy-stability monitor and a payments team's fraud kill-switch require different implementation ownership and, per Section~\ref{sec:payreg}, are bound by different external clocks. A firm should not treat completion of the trading track as a precondition for beginning this one; the inventory step in Days~1--30 of each track can and should run concurrently since the same foundation-model vendor may appear in both a trading pipeline's Layer~3 disclosure (Section~\ref{sec:l25msrr}) and a fraud pipeline's extended attestation (Section~\ref{sec:l25fraud}), and a firm that discovers this overlap only at Day~90 of one track has lost the opportunity to attest to it jointly.

\section{Discussion: The Literacy Problem Underneath the Governance Gap}\label{sec:literacy}
The survey evidence in Section~\ref{sec:survey} points to an institutional failure that no framework can resolve on its own: the people responsible for governing these systems do not yet possess the technical literacy to specify governance for them. The news-anomaly case study sharpens this point: literacy sufficient to govern an in-house RL trading agent is not sufficient to govern a strategy where the risk profile depends on a foundation model that an institution licenses but does not build, evaluated against a news feed shared with competitors.

\section{Limitations}\label{sec:limits}
The survey evidence is self-selected and likely understates the true governance gap. The adversarial-input protocol (Section~\ref{sec:adversarial}) and the systemic-layer simulation (Section~\ref{sec:l3calib}) remain illustrative rather than fully empirically populated. The taxonomy and ARS are a triage instrument, not a validated risk metric. The news-anomaly case study in Section~\ref{sec:msrr} is drawn from a single strategy family and asset class (U.S.\ equities, Reuters/Dow Jones news, 1996--2022). The magnitude of vendor-dependency and crowding effects documented there should not be read as a calibrated estimate of these channels for other agentic strategy classes, though the qualitative mechanisms (opaque vendor-model dependency, undisclosed vendor-side updates, and shared-feed crowding) generalize directly.

The Layer~2 drift monitor has limits of its own. The integrated-covariance identity is exact only for decisions produced by an optimizer of an objective linear in the cost, with the remaining terms fixed over the evaluation window; the replay integrity check detects some, but not all, departures from this premise. The exact tier needs replay access to the decision layer, which vendor contracts do not yet routinely grant, and it is not cheaper than re-solving the decision problem scenario by scenario; its value is that it is exact and explanatory, not that it saves computation \citep{aldridge2026}. The monitor measures departure from the validated policy, so it inherits any flaw in that policy and, through the loss function, any mispricing in the reviewed cost matrix. In the fraud domain it needs realized labels, which arrive on the chargeback lag, and its power against weak drift is limited (Appendix~\ref{app:robust}). Both worked examples are synthetic, with a single abrupt change point, and have not been validated on production data.

\section{Conclusion}\label{sec:concl}
The governance gap documented in Section~\ref{sec:gap} is an architectural literacy gap, not a compliance gap. Across four failure modes (endogenous policy drift, adversarial input-distribution drift, compositional risk, and cross-institutional policy correlation), frameworks built for deterministic systems do not transfer to agentic systems. The news-anomaly case study of Section~\ref{sec:msrr} shows these are not abstract future risks: a currently deployed, economically large, vendor-dependent, periodically retrained agentic strategy already exhibits compositional fragility to vendor model updates and a market-level crowding signature consistent with our Layer~4 mechanism. For the first of these failure modes, the integrated-covariance identity gives the engineering layer an exact, computable footing: regret is the covariance between costs and decisions accumulated over the level of uncertainty, the plain covariance is a conservative bound available from observed data alone, and replay access to the decision layer, not access to model weights, is what makes the measurement exact. Institutions that build governance capability for systems of this kind now will not be retrofitting it under regulatory pressure in 2027.

\appendix
\section{Appendix: Robustness of the Integrated-Covariance Monitor}\label{app:robust}
Sections~\ref{sec:l2drift} and~\ref{sec:l2fraud} each report detection statistics from a single calibration ($W=20$, one random seed) to keep the worked examples legible. This appendix reports the same monitors' behavior across a grid of $(W,\theta,k)$ and 30 independent random seeds per configuration, to establish that the headline detection latencies in Tables~\ref{tab:trading} and~\ref{tab:fraud} are representative of the estimators' behavior rather than an artifact of a favorably drawn sample path.

\paragraph{Protocol.} For each configuration, we generate 30 independent synthetic environments (distinct random seeds), each following the change-point construction of Section~\ref{sec:l2drift} (trading: reactivity rising from $\theta_0=0.2$ to $\theta$ in \eqref{eq:tradingrule}) or Section~\ref{sec:l2fraud} (fraud: a spurious segment loading $\theta$ in \eqref{eq:fraudrule}, with the decline rate recalibrated): $T=500$ periods, a validated regime for $t<300$, and a drifted regime for $t\geq300$. Within each trial, we calibrate $\tau$ on an early sub-window of the validated regime ($t\in[W,200)$) and hold out the remainder of the validated regime ($t\in[200,300)$) purely for out-of-sample false-alarm measurement, so that the reported false-alarm rate is not measured against the same data used to set the trigger level. The statistic is the replay-tier integrated-covariance regret in the trading domain (with the passive half-covariance reported for comparison) and the regret-to-covariance ratio $\hat\rho_t$ in the fraud domain. We report, across the 30 seeds: the fraction of trials in which the monitor fires during the 200 drifted periods (detection rate), the mean and standard deviation of detection latency conditional on firing, and the mean out-of-sample false-alarm rate during the held-out validated window. The grid is the full cross product $W\in\{10,20,30\}$, $\theta\in\{0.8,1.2,1.6\}$, $k\in\{3,4,5\}$: 27 configurations and 810 trials per domain.

\paragraph{Result: trading.} The replay monitor detected the drift in all 810 trials; no configuration produced a missed detection. The passive monitor detected it in 99.9\% of trials (lowest configuration: 97\%) but with mean latencies roughly three times longer. Table~\ref{tab:rob_trading} reports the marginal sensitivity to each parameter, holding the other two at the main-text values ($W=20$, $\theta=1.2$, $k=4$).

\begin{table}[ht]
\centering\small
\caption{Trading domain: marginal parameter sensitivity, 30 seeds per row. Latency in periods. The false-alarm rate does not depend on $\theta$, which affects only the drifted regime.}\label{tab:rob_trading}
\begin{tabular}{llccccc}
\toprule
& & \multicolumn{4}{c}{Replay tier (integrated covariance)} & Passive tier\\
\cmidrule(lr){3-6}\cmidrule(lr){7-7}
Varying & Value & Detection & Mean lat. & Std.\ lat. & False-alarm rate & Mean lat.\\
\midrule
$W$ (fix $\theta=1.2$, $k=4$) & 10 & 100\% & 5.2 & 4.3 & 0.0063 & 16.9\\
& 20 & 100\% & 5.5 & 5.1 & 0.0097 & 16.3\\
& 30 & 100\% & 6.9 & 5.4 & 0.0127 & 18.3\\
\addlinespace
$\theta$ (fix $W=20$, $k=4$) & 0.8 & 100\% & 10.6 & 9.9 & 0.0097 & 30.9\\
& 1.2 & 100\% & 5.5 & 5.1 & 0.0097 & 16.3\\
& 1.6 & 100\% & 4.4 & 4.0 & 0.0097 & 14.3\\
\addlinespace
$k$ (fix $W=20$, $\theta=1.2$) & 3 & 100\% & 4.3 & 4.0 & 0.0233 & 12.5\\
& 4 & 100\% & 5.5 & 5.1 & 0.0097 & 16.3\\
& 5 & 100\% & 8.3 & 8.8 & 0.0027 & 25.3\\
\bottomrule
\end{tabular}
\end{table}

\paragraph{Result: fraud.} The $\hat\rho_t$ monitor detected every drift of strength $\theta\geq1.2$ at $k=3$, for every window length, but its power falls for weak drift and for high $k$ (Table~\ref{tab:rob_fraud_det}); across all 810 trials it detected 79\% of drifts. Table~\ref{tab:rob_fraud} reports the marginal sensitivity around the main-text values ($W=20$, $\theta=1.2$, $k=3$).

\begin{table}[ht]
\centering\small
\caption{Fraud domain: marginal parameter sensitivity of the $\hat\rho_t$ monitor, 30 seeds per row. Latency in batches, conditional on detection within 200 batches.}\label{tab:rob_fraud}
\begin{tabular}{llcccc}
\toprule
Varying & Value & Detection rate & Mean latency & Std.\ latency & Mean false-alarm rate\\
\midrule
$W$ (fix $\theta=1.2$, $k=3$) & 10 & 100\% & 16.6 & 11.4 & 0.0013\\
& 20 & 100\% & 17.9 & 11.2 & 0.0043\\
& 30 & 100\% & 17.9 & 8.8 & 0.0077\\
\addlinespace
$\theta$ (fix $W=20$, $k=3$) & 0.8 & 83\% & 65.4 & 54.0 & 0.0043\\
& 1.2 & 100\% & 17.9 & 11.2 & 0.0043\\
& 1.6 & 100\% & 9.0 & 4.0 & 0.0043\\
\addlinespace
$k$ (fix $W=20$, $\theta=1.2$) & 3 & 100\% & 17.9 & 11.2 & 0.0043\\
& 4 & 100\% & 35.1 & 30.5 & 0.0000\\
& 5 & 80\% & 82.7 & 60.1 & 0.0000\\
\bottomrule
\end{tabular}
\end{table}

\begin{table}[ht]
\centering\small
\caption{Fraud domain: detection rate of the $\hat\rho_t$ monitor over the full grid, 30 seeds per cell. Columns give the window length $W$ (batches).}\label{tab:rob_fraud_det}
\begin{tabular}{llccc}
\toprule
Drift strength & Trigger & $W=10$ & $W=20$ & $W=30$\\
\midrule
$\theta=0.8$ & $k=3$ & 67\% & 83\% & 93\%\\
& $k=4$ & 30\% & 43\% & 57\%\\
& $k=5$ & 7\% & 13\% & 33\%\\
\addlinespace
$\theta=1.2$ & $k=3$ & 100\% & 100\% & 100\%\\
& $k=4$ & 83\% & 100\% & 100\%\\
& $k=5$ & 47\% & 80\% & 93\%\\
\addlinespace
$\theta=1.6$ & $k=3$ & 100\% & 100\% & 100\%\\
& $k=4$ & 100\% & 100\% & 100\%\\
& $k=5$ & 97\% & 100\% & 100\%\\
\bottomrule
\end{tabular}
\end{table}

\paragraph{Interpretation.} Four patterns follow from the estimators' construction rather than being empirical surprises: (i) detection latency falls monotonically as $\theta$ increases, since stronger drift produces a larger shift in the cost--decision covariance profile; (ii) latency rises monotonically with $k$, the direct consequence of demanding more standard deviations of evidence before alarming, while the false-alarm rate, already low, falls further; (iii) $W$ has a comparatively small effect on latency but raises the false-alarm rate in both domains, consistent with a longer trailing window producing a smoother but more persistent statistic that, once it drifts above threshold through estimation noise, stays there longer; and (iv) in the trading domain, the replay tier is about three times faster than the passive tier at every configuration, because replay compares each decision with its own counterfactual and so removes the noise contributed by other inputs. Two further points are specific to the fraud domain. First, these results are the quantitative basis for the $k=3$ (fraud) versus $k=4$ (trading) choice argued in Section~\ref{sec:l2fraud}: in the fraud domain, raising $k$ from 3 to 4 doubles mean latency and raising it to 5 misses one drift in five, while $k=3$ keeps the out-of-sample false-alarm rate below 1\% for every window length, a trade an institution facing disparate-impact exposure would reasonably take and one a trading desk facing only a costly-but-non-regulatory review need not. Second, against weak drift ($\theta=0.8$) the fraud monitor's power is limited at any $k$; longer windows help, at the cost of a higher false-alarm rate.

\paragraph{Scope of this robustness check.} This grid varies the monitors' own parameters and the drift strength against a fixed data-generating process (a single abrupt step change at a known $t=300$). It does not test robustness to misspecification of the change type itself, e.g., a gradual drift rather than a step change, an intermittent or regime-switching drift, or a change in the noise variance rather than in the decision rule. We do not claim the parameter choices reported in Tables~\ref{tab:rob_trading}--\ref{tab:rob_fraud_det} are optimal in any formal sense; they establish how detection responds to reasonable perturbation of $(W,\theta,k)$ around the values used in the main text, which is the narrower and more defensible claim this appendix is designed to support. An adopting institution should re-run this grid, or a finer one, against its own historical cost/decision or chargeback/decline series before selecting production values.

\end{document}